\documentclass{LMCS}
\def\doi{8 (3:13) 2012}
\lmcsheading%
{\doi}
{1--30}
{}
{}
{Jan.~17, 2011}
{Sep.~12, 2012}
{}

\usepackage{graphicx}
\usepackage{color}
\usepackage{mathrsfs}         
\usepackage{enumerate}

\newcommand{\Inv}[0]{\ensuremath{\mathrm{Inv}}}
\newcommand{\Pol}[0]{\ensuremath{\mathrm{Pol}}}
\newcommand{\Aut}[0]{\ensuremath{\mathrm{Aut}}}
\newcommand{\End}[0]{\ensuremath{\mathrm{End}}}

\renewcommand{\phi}{\varphi}

\newcommand{\ignore}[1]{}
\newcommand{\A}{\mathfrak{A}}
\newcommand{\B}{\mathfrak{B}}
\newcommand{\C}{\mathfrak{C}}

\newcommand{\M}{\mathfrak{M}}
\newcommand{\bA}{\mathfrak{A}}
\newcommand{\bB}{\mathfrak{B}}
\newcommand{\bC}{\mathfrak{C}}
\newcommand{\bD}{\mathfrak{D}}
\newcommand{\bE}{\mathfrak{E}}

\begin{document}

\title[The Universal-Algebraic Approach to Constraint Satisfaction]{On the Scope of the Universal-Algebraic Approach to Constraint Satisfaction\rsuper*}

\author[M.~Bodirsky]{Manuel Bodirsky\rsuper a}
\address{{\lsuper a}CNRS/LIX, \'Ecole Polytechnique, France}
\email{bodirsky@lix.polytechnique.fr}

\author[M.~Hils]{Martin Hils\rsuper b}
\address{{\lsuper b}Institut de Math\'ematiques de Jussieu, Universit\'e Paris Diderot - Paris 7, France}
\email{hils@logique.jussieu.fr}

\author[B.~Martin]{Barnaby Martin\rsuper c}
\address{{\lsuper c}Engineering and Computing Sciences, Durham University, U.K.}
\email{barnabymartin@gmail.com}

\thanks{The first author has received funding from the European Research Council under the European Community's Seventh Framework Programme (FP7/2007-2013 Grant Agreement no. 257039). The third author was supported by EPSRC grant EP/G020604}
\titlecomment{{\lsuper*}Many of the results of this paper have appeared in the extended abstract \cite{BodHilsMartin}.}

\begin{abstract}
The universal-algebraic approach has proved a powerful tool in the study
of the computational complexity of constraint satisfaction problems (CSPs).
This approach has previously been applied to the study of CSPs with
finite or (infinite) $\omega$-categorical templates,
and relies on two facts. The first is that
in finite or $\omega$-categorical structures $\A$, a relation
is primitive positive definable if and only if it is preserved by the polymorphisms
of $\A$. The second is that every finite or $\omega$-categorical
 structure is homomorphically equivalent to a core structure.

In this paper, we present generalizations of these facts to infinite
structures that are not necessarily $\omega$-categorical.
Specifically, we prove that every CSP can be formulated with a template $\A$ such that
a relation is primitive positive definable in $\A$ if and only if it is first-order definable in
$\A$ and preserved by the \emph{infinitary} polymorphisms of $\A$.
Using existential positive closure we rederive and extend known results about cores, presenting the new notions of core theories (models of which will be cores) and core companions (which are
defined analogously to model companions in model theory).
We prove a uniqueness result for core companions that yields the uniqueness
of model-complete cores of $\omega$-categorical structures.
Existential positive closure is also the crucial concept to
give an exact characterization of those CSPs that
can be formulated with (a finite or) an $\omega$-categorical template.

Finally, we present applications of our general results to the description
and analysis of the computational complexity of CSPs. In particular,
we give general hardness criteria based on the absence of polymorphisms
that depend on more than one argument, and we present a polymorphism-based description of those CSPs that
are first-order definable (and therefore can be solved in polynomial time).
\end{abstract}

\maketitle

\section{Introduction}
For a relational structure $\A$ over a finite signature the \emph{constraint satisfaction problem} CSP$(\A)$ is the computational problem to decide whether a primitive positive first-order sentence $\phi$ -- that is, the existential quantification of a conjunction of atomic formulas -- is true on $\A$. The case where the template $\A$ is finite has been extensively studied in the literature, and is known to comprise a significant microcosm of the complexity class NP (see, e.g., \cite{Jeavons}). 
The universal-algebraic approach, of studying the invariance properties of relations under the action of polymorphisms, has been particularly powerful in the complexity analysis of finite-domain CSPs (see \cite{JeavonsClosure} as a starting point). This approach has also been successfully used in infinite-domain CSPs where the template is \emph{$\omega$-categorical}, i.e., is the unique countably infinite model of its first-order (fo) theory up to isomorphism -- see, e.g., \cite{tcsps}.

Many interesting problems can be formulated as infinite CSPs whose template is not $\omega$-categori\-cal. To illustrate the wealth of the class of CSPs studied in this paper, we present three concrete computational problems that can be formulated as CSP$(\A)$, for an infinite $\A$. Each of these problems is solvable in polynomial time -- and the proofs of this are generally non-trivial. The templates $(\mathbb{Z};+,1)$ and $(\mathbb{R};+,1)$, where $+$ is read as the ternary relation $x+y=z$, correspond to the solving of Linear Diophantine Equations and Linear Real Equations, respectively. 
Another template of interest relates to the Unification Problem. Let $\sigma:=(f_1,f_2,\ldots)$ be a functional signature, we form the template $(T;F_1,F_2,\ldots)$, where $T$ is the term algebra on $\sigma$ built over a countably infinite set of variables, and each $F_i$ is the relational form $f_i(t_1,\ldots,t_{r_i})=t_0$ of $f_i$ over $T$. 

\paragraph{\textbf{Primitive positive definability of formulas}.} A central role is played in CSPs by the notion of primitive positive (pp) definability. Two templates give rise to the same CSP iff they agree on all pp-sentences (have the same pp-theory). A less central, but often equivalent role is played by existential positive (ep) definability. For example, two templates have the same pp-theory iff they have the same ep-theory. If $\langle \A \rangle_{\textrm{pp}}$ are the relations pp-definable on $\A$, then $\langle \A \rangle_{\textrm{pp}} \subseteq \langle \A' \rangle_{\textrm{pp}}$ implies that there is a polynomial time reduction from $\mathrm{CSP}(\A')$ to $\mathrm{CSP}(\A)$. In the case that $\A$ is finite or $\omega$-categorical, the set of relations $\langle \A \rangle_{\textrm{pp}}$ are precisely those that are invariant under the polymorphisms of $\A$ -- $\Inv(\Pol(\A))$. We note that this relationship holds on some infinite structures which are not $\omega$-categorical (an example is given in \cite{RomovPosPrimStruc}, also the natural numbers with the binary successor relation may easily be verified to have this property). In general, a CSP can not be formulated with (a finite or) an $\omega$-categorical template (we return to this below). However, we are able to prove the existence of an equivalent template -- with the same pp-theory -- which enjoys some of the benign properties of finiteness or $\omega$-categoricity. Given any $\A$, we prove the existence of a highly saturated elementary extension $\M$ such that a relation is pp-definable on $\M$ iff it is fo-definable on $\M$ and invariant under the polymorphisms of (countably) infinite arity of $\M$. In fact, we prove that this relationship holds for all saturated structures of cardinality at least $2^\omega$. But our construction obviates the need for the set-theoretic assumptions usually required to assert the existence of a saturated elementary extension of an arbitrary structure. However, in many concrete cases, such as for structures that are uncountably categorical, such saturated models arise directly. 
We go on to prove that each of the three assumptions -- high saturation, infinitary (and not finitary) polymorphism and fo-intersection -- is necessary. That is, we exhibit structures for which any two of these is insufficient for the respective connection. 

\paragraph{\textbf{Existential-positively closed models}.} We use existential-positively closed (epc) models to easily rederive and extend known results in the literature from \cite{Cores-journal}. In the case of finite-domain structures there is the well-established notion of core -- a structure, homomorphically equivalent to the original, all of whose endomorphisms are embeddings (equivalently, in the finite, automorphisms). The core is unique up to isomorphism, and the task of problem classification is simplified as one may restrict attention to cores. For $\omega$-categorical structures we no longer have uniqueness of cores in general, but it was shown in~\cite{Cores-journal} that every $\omega$-categorical structure is homomorphically equivalent to a model-complete core, i.e. to a structure whose endomorphisms preserve all first-order formulas, that is unique up to isomorphism and again $\omega$-categorical. We present a new proof here, on the way introducing the concepts of core theories, models of which will be cores, and core companions. If an ep-theory $S$ is contained in 
a model-complete core theory $T$ -- with the same existential positive and universal negative restriction -- which will be called the \emph{core companion} of $S$, then $T$ is unique up to equivalence of theories. In the special case of $\omega$-categorical structures, this yields the uniqueness result for model-complete cores. We give a necessary and sufficient condition as to when CSP$(\A)$ can be formulated with a template which has a model-complete core theory, based on whether the class of epc models of a restricted part of $\A$'s $\forall \exists$-theory is axiomatizable. 

When distinct templates give rise to the same CSP, it might be the case that one is better behaved than another. For example, $(\mathbb{Z}; <)$ and $(\mathbb{Q}; <)$ share the same ep-theory (pp-theory); yet while $\Inv(\Pol(\mathbb{Z}; <)) \neq \langle (\mathbb{Z}; <) \rangle_{\textrm{pp}}$, we have $\Inv(\Pol(\mathbb{Q}; <)) = \langle (\mathbb{Q}; <) \rangle_{\textrm{pp}}$, as $(\mathbb{Q}; <)$ is $\omega$-categorical. In the present paper we give a necessary and sufficient condition that a template $\A$ has an equivalent $\A'$ that is finite or $\omega$-categorical -- and consequently satisfies $\Inv(\Pol(\A'))=\langle \A' \rangle_{\textrm{pp}}$. The condition is that the number of maximal ep-$n$-types consistent with the theory of $\A$ is finite, for all $n$. It follows that none of the three examples of the previous page may be formulated with an $\omega$-categorical template. Our result uses epc models and the direct limit construction.
 
\paragraph{\textbf{Applications}.} We go on to consider the repercussions of our restricted relationship for the complexity of CSPs. Firstly, we show that existential positive (ep) and pp-definability coincide on a structure $\A$ iff all $\omega$-polymorphisms of all elementary extensions of $\A$ are essentially unary. We demonstrate that the move to elementary extensions is necessary by giving a structure whose $\omega$-polymorphisms include only projections but for which $(x=y \vee u=v)$ is not pp-definable. Using the notion of local refutability in \cite{BodirskyHermannRichoux}, we note that if a structure $\A$ is not locally refutable, and all $\omega$-polymorphisms of all elementary extensions of $\A$ are essentially unary, then CSP$(\A)$ is NP-hard. 

Secondly, and introducing our philosophy to the work of \cite{LLT}, we present a polymorphism based description of those CSPs that are fo-definable. We show that CSP$(\mathfrak A)$ is fo-definable if and only if $\mathfrak A$ has an elementary extension which has a $1$-tolerant polymorphism. It follows that such CSPs are polynomial-time solvable. 

Thirdly, we recall a known relationship between certain binary injective polymorphisms and Horn definability (given in the context of $\omega$-categorical structures in \cite{Maximal}). Considering as a polymorphism an embedding $e$ of $(\mathbb{R};+,1)^2$ into $(\mathbb{R};+,1)$, we show that the recent complexity classification of \cite{HornOrFull} may be given a natural algebraic specification. The presence of the polymorphism $e$ separates those fo-expansions of $(\mathbb{R};+,1)$ whose CSP is in P from those whose CSP is NP-complete. Thus we demonstrate that the presence of certain polymorphisms can delineate complexity even outside of the realm of $\omega$-categoricity.

\paragraph{\textbf{Related work}.} There are several extant works on notions of pp-definability over infinite structures, including those involving infinitary polymorphisms and infinitary relations \cite{SzaboLoc,Poizat:AlgebresDePost,RomovISMVL04}. Relational operations transcending normal pp-definitions are usually permitted, such as: infinite conjunction, infinite projection and various forms of monotone disjunction. In order for our results to be applicable to the (finite!) instances of CSPs, we are not able to sacrifice anything on the relational side, and so pp-definability must remain in its most basic form. This represents the principle difference between our work and those that have come before.  We note that this is the first time that infinitary polymorphisms have been considered in connection with the complexity of CSPs.

An extended abstract of this paper appeared at conference as \cite{BodHilsMartin}. Section~\ref{sec:core-theories}, as well as many previously omitted proofs, appear here for the first time.

\section{Preliminaries}
\subsection{Models, operations and theories}
A \emph{relational signature} (with constants) $\tau$ is a set of relation symbols
$R_i$, each of which has an associated finite arity $k_i$, and constants $c_i$. We consider only relational signatures (with constants) in this paper.
A (relational) {\em structure} $\A$ over the signature $\tau$ (also called $\tau$-\emph{structure}) consists of a set $A$ (the {\em domain}) together with a relation $R^\A \subseteq A^k$ for each relation symbol $R$ of arity $k$ from $\tau$ and a constant $c^\A \in A$ for each constant symbol $c$. 

Let $\A$ be a $\tau$-structure, and let $\A'$ be a
$\tau'$-structure with $\tau \subseteq \tau'$. If
$\A$ and $\A'$ have the same domain and 
$R^\A = R^{\A'}, c^\A=c^{\A'}$ for all $R,c \in \tau$, then $\A$
is called the \emph{$\tau$-reduct} (or simply \emph{reduct}) of $\A'$,
and $\A'$ is called a \emph{$\tau'$-expansion} (or simply \emph{expansion}) of $\A$.
If $\A$ is a $\tau$-structure and $\langle a_\alpha \rangle_{\alpha <\beta}$ is a sequence of elements of $A$, then $(\A;\langle a_\alpha \rangle_{\alpha <\beta})$ is the natural $\tau \cup \{c_\alpha : \alpha <\beta\}$-expansion of $\A$ with $\beta$ new constants, where $c_\alpha$ is interpreted by $a_\alpha$. $\A$ is an \emph{extension} of $\B$, denoted $\B \subseteq \A$, if $B \subseteq A$ and for each $R$ in $\tau$, and for all tuples $\overline{b}$ from $B$, $\overline{b} \in R^\B$ iff $\overline{b} \in R^\A$, and for each $c$ in $\tau$, $c^\B=c^\A$. Let $\langle b_\alpha \rangle_{\alpha <|B|}$ well-order the elements of $\B$. $\A$ is an \emph{elementary extension} of $\B$, denoted $\B \preceq \A$, if it is an extension and, for each first-order (fo) $\tau \cup \{c_\alpha : \alpha <|B|\}$-sentence $\phi$, $(\B,\langle b_\alpha \rangle_{\alpha <|B|}) \models \phi$ iff $(\A,\langle b_\alpha \rangle_{\alpha <|B|}) \models \phi$.  

An fo-formula is \emph{existential positive} (ep) if it involves only existential quantification, conjunction and disjunction (no negation). Furthermore, if it involves no instances of disjunction, then it is termed \emph{primitive positive} (pp). Note that we consider the boolean false $\bot$ to be a pp-formula, and we always allow equalities in pp-formulas. Dual to ep are \emph{universal negative} formulas, involving only universal quantification, conjunction and disjunction, with all atoms negated.
Suppose $\A$ is a finite structure over a finite signature with domain $A:=\{a_1,\ldots,a_s\}$. Let $\theta(x_1,\ldots,x_s)$ be the conjunction of the positive facts of $\A$, where the variables $x_1,\ldots,x_s$ correspond to the elements $a_1,\ldots,a_s$. That is, $R(x_{\lambda_1},\ldots,x_{\lambda_k})$ appears as an atom in $\theta$ iff $(a_{\lambda_1},\ldots,a_{\lambda_k}) \in R^\A$. Define the \emph{canonical query} $\phi[\A]$ of $\A$ to be the pp-formula $\exists x_1 \ldots x_s. \theta(x_1,\ldots,x_s)$.
%
A set of formulas $\Phi:=\Phi(x_1,\ldots,x_n)$ with free variables $x_1,\dots,x_n$ is called \emph{satisfiable in $\A$} if there are elements $a_1,\dots,a_n$ from $A$ such that for all sentences $\phi \in \Phi$ we have $\A \models \phi(a_1,\dots,a_n)$.
We say that $\Phi$ is \emph{satisfiable} if there exists a structure
$\A$ such that $\Phi$ is satisfiable in $\A$.

A \emph{$\tau$-theory} is a set of $\tau$-sentences; two theories are \emph{equivalent} if they share the same models. For a $\tau$-structure $\A$, define the \emph{theory} of $\A$, $\mathrm{Th}(\A)$, to be the set of $\tau$-sentences true on $\A$. Note that $\A \preceq \B$ implies that $\mathrm{Th}(\A)=$ $\mathrm{Th}(\B)$. Define $\mathrm{Th}_{\forall^-}(\A)$ to be the set of universal negative sentences true on $\A$.

For $n \geq 0$, an \emph{$n$-type} of a theory $T$ is a set $p:=p(x_1,\ldots,x_n)$ of formulas in the free variables $x_1,\dots,x_n$ such that $p \cup T$ is satisfiable. In a similar manner, a \emph{primitive positive $n$-type} (pp-$n$-type) of a theory $T$ is a set of pp-formulas such that $p \cup T$ is satisfiable. A pp-$n$-type $p$ of $T$ is \emph{maximal} if $T \cup p \cup \phi(x_1,\ldots,x_n)$ is unsatisfiable for any pp-formula $\phi \notin T \cup p$.  
A (pp-) $n$-type of a structure $\A$ is just a (pp-) $n$-type of the theory $\mathrm{Th}(\A)$. 
%
We define (maximal) ep-$n$-types in exactly the like fashion, with ep-formulas. A \emph{complete} ep-$n$-type of a structure $\bA$ is the set of all ep-formulas holding on some tuple from $\bA$.
An $n$-type $p(x_1,\ldots,x_n)$ of $\A$ is \emph{realized} in $\A$ if there exists $a'_1,\ldots,a'_n \in A$ s.t., for each $\phi \in p$, $\A \models \phi(a'_1,\ldots,a'_n)$. For an infinite cardinal $\kappa$, a structure $\A$ is \emph{$\kappa$-saturated} if, for all $\beta < \kappa$ and expansions $(\A;\langle a_\alpha\rangle_{\alpha <\beta})$ of $\A$, every $1$-type of $(\A;\langle a_\alpha\rangle_{\alpha <\beta})$ is realized in $(\A;\langle a_\alpha\rangle_{\alpha <\beta})$. We say that an infinite $\A$ is \emph{saturated} when it is $|A|$-saturated. Realization of pp- (ep-)types and pp- (ep-)($\kappa$-)saturation is defined in exactly the analogous way. Note that a structure that is $\kappa$-saturated is a fortiori ep-$\kappa$-saturated. (Ep-$\kappa$-saturation and pp-$\kappa$-saturation in fact coincide, though we will only need to apply the trivial direction of this relationship.) A theory $T$ is said to be \emph{$\kappa$-categorical}, for some cardinal $\kappa$, if it has a unique model of cardinality $\kappa$, up to isomorphism. It is known that, if $T$ is $\kappa$-categorical for one uncountable cardinal $\kappa$, then $T$ is $\kappa'$-categorical for all uncountable cardinals $\kappa'$. A structure $\A$, of cardinality $\kappa$, is said to be $\kappa$-categorical if $\mathrm{Th}(\A)$ is $\kappa$-categorical.

Let $\A$ and $\B$ be $\tau$-structures.  A \emph{homomorphism} from
$\A$ to $\B$ is a function $f$ from $A$ to $B$
such that for each $k$-ary relation symbol $R$ in $\tau$ and each $k$-tuple
$(a_1, \dots, a_k)$, if $(a_1, \dots, a_k) \in R^\A$, then $(f(a_1),
\dots, f(a_k)) \in R^\B$. Also, for each constant $c$ in $\tau$, $f(c^\A)=c^\B$. In this case we say that the map $f$
\emph{preserves} the relation $R$.  
Injective homomorphisms that also preserve
the complement of each relation are called \emph{embeddings}.  Surjective
embeddings are called isomorphisms; homomorphisms and isomorphisms from $\A$ to itself are called \emph{endomorphisms} and \emph{automorphisms}, respectively. An \emph{elementary embedding} is an embedding that preserves all fo-formulas. A structure is a \emph{core} if all its endomorphisms are embeddings \cite{Cores-journal} (in the finite this coincides with the more usual definition of all endomorphisms being automorphisms). $\B$ is \emph{a core} of $\A$ if $\B$ is a core and $\A$ and $\B$ are homomorphically equivalent. 
We will make use later of the following lemma, a close relative of Theorem 10.3.1 in \cite{HodgesLong}.
\begin{lem}\label{lem:sat-hom}
Let $\A$ and $\B$ be $\tau$-structures, where $\B$ is pp-$|A|$-saturated. Suppose $f$ is a mapping from $\{a_\alpha : \alpha < \mu\} \subseteq A$ ($\mu < |A|$) to $B$ such that
all pp-($\tau \cup \{c_\alpha : \alpha<\mu\}$)-sentences true on $(\A;\langle a_\alpha \rangle_{\alpha<\mu})$ are true on $(\B;\langle f(a_\alpha) \rangle_{\alpha<\mu})$. Then $f$ can be extended to a homomorphism from $\A$ to $\B$.
\end{lem}
\proof
Note that, if $\B$ is finite, then $\B$ is pp-$|A|$-saturated no matter what cardinality $\A$ is.


Suppose $\mu < |A|=\kappa_A$. Let $\langle a'_\alpha \rangle_{\alpha < \kappa_A}$ well-order $\A$ such that $\langle a'_\alpha \rangle_{\alpha<\mu}=$ $\langle a_\alpha \rangle_{\alpha<\mu}$ (there is the implicit and harmless assumption that $\langle a_\alpha \rangle_{\alpha<\mu}$ contains no repetitions). Set $\langle b'_\alpha \rangle_{\alpha<\mu}:=$ $\langle f(a_\alpha) \rangle_{\alpha<\mu}$. 

We will construct by transfinite recursion on $\beta$ (up to $\kappa_A$) a sequence $\langle b'_\alpha\rangle_{\alpha<\beta}$ such that we maintain the inductive hypothesis
\begin{iteMize}{$(*)$}
\item all pp-($\tau \cup \{c_\alpha : \alpha<\beta\}$)-sentences true on $(\A;\langle a'_\alpha\rangle_{\alpha<\beta})$ are true on $(\B;\langle b'_\alpha\rangle_{\alpha<\beta})$.
\end{iteMize}
The result will clearly then follow by reading $f$ as the map $\{a'_\alpha \mapsto b'_\alpha\}_{\alpha< \kappa_A}$.

(Base Case.) $\beta:=\mu$. Follows from hypothesis of lemma.

(Inductive Step. Limit ordinals.) $\beta:=\lambda$. Property $(*)$ holds as a sentence can only mention a finite collection of constants, whose indices must all be less than some $\gamma < \lambda$. 

(Inductive Step. Successor ordinals.) $\beta:=\gamma+1 < \kappa_A$. Set
\[
\begin{array}{l}
\Sigma := \{ \phi(x) : \phi \mbox{ is a pp-($\tau \cup \{c_\alpha:\alpha<\gamma\}$)-formula s.t.} (\A;\langle a'_\alpha \rangle_{\alpha<\gamma}) \models \phi(a'_\gamma) \}.
\end{array}
\] 
By $(*)$, for every $\phi \in \Sigma$, $(\B;\langle b'_\alpha \rangle_{\alpha<\gamma}) \models \exists x. \phi(x)$. By compactness, since $\Sigma$ is closed under conjunction, we have that $\Sigma$ is a pp-$1$-type of $(\B;\langle b'_\alpha \rangle_{\alpha<\gamma})$. By pp-$|A|$-saturation of $\B$ it is realized by some element $b'_\gamma \in B$. By construction we maintain that all pp-($\tau \cup \{c_\alpha : \alpha<\gamma+1\}$)-sentences true on $(\A;\langle a'_\alpha\rangle_{\alpha<\gamma+1})$ are true on $(\B;\langle b'_\alpha\rangle_{\alpha<\gamma+1})$.
\qed
\noindent The previous lemma, as so often in this paper, holds equally for ep, in place of pp. However, we will want to apply it later to specifically the primitive positive. As a rule, we choose ep over pp in all cases where either could be used without affecting the exposition.

For a sequence of $\tau$-structures $\A_\alpha$, $\alpha<\mu$, define the \emph{direct} (or categorical) product $\prod_{\alpha < \mu} \A_\alpha$ to be the $\tau$-structure on domain $\prod_{\alpha < \mu} A_\alpha$
such that $(\langle a^1_\alpha \rangle_{\alpha<\mu},\ldots,\langle a^r_\alpha\rangle_{\alpha<\mu}) \in R^{\prod_{\alpha < \mu} \A_\alpha}$ iff $(a^1_\alpha,\ldots,a^r_\alpha) \in R^{\A_\alpha}$ for each $\alpha < \mu$. Also, $c^{\prod_{\alpha < \mu} \A_\alpha}=\langle c^{\A_\alpha} \rangle_{\alpha<\mu}$.
Note that short direct products are indicated infix with $\times$. A property of pp-sentences $\phi$ that we will use later is that $\A \models \phi$ and $\B \models \phi$ iff $\A \times \B \models \phi$.

Let $\langle \A\rangle_{\mathrm{fo}}$ (respectively, $\langle \A\rangle_{\mathrm{ep}}$ and $\langle \A\rangle_{\mathrm{pp}}$) be the sets of relations, over domain $A$, that are fo- (respectively, ep- and pp-) definable over $\A$ (without parameters). 
Let $\Aut(\A)$ and $\End(\A)$ be the sets of automorphisms and endomorphisms, respectively, of $\A$.
A \emph{$\kappa$-polymorphism} of $\A$ is a homomorphism from $\A^\kappa$ to $\A$, where the power is with respect to the direct product already defined. Let $\Pol^\infty(\A)$, $\Pol^\omega(\A)$ and $\Pol(\A)$ be the sets of $\kappa$-polymorphisms (for any $\kappa$), $\kappa$-polymorphisms (for $\kappa \leq \omega$) and $k$-polymorphisms (for each finite $k$), respectively. For a set of operations $F$ on domain $A$, define $\Inv(F)$ to be the set of relations, over $A$, that are preserved by (invariant under) each of the operations in $F$ (note that the condition of preservation of an $m$-ary relation by a $\kappa$-ary function $f:\A^\kappa \rightarrow \A$ is \emph{component-wise}, i.e. if $(a^\beta_1,\ldots,a^\beta_m) \in R^\A$, for all $\beta < \kappa$, then $(f(\langle a^\beta_1\rangle_{\beta < \kappa}), \ldots,f(\langle a^\beta_m\rangle_{\beta < \kappa})) \in R^\A$).

Let $\overline{t}$ be a $k$-tuple of elements from a structure $\A$. Then the \emph{orbit} of $\overline{t}$ under Aut$(\A)$
is the set $\{ f(\overline{t}) \; | \; f \in \text{Aut}(\A) \}$.
\begin{thm}[Ryll-Nardzewski, Engeler, Svenonius; see e.g.~\cite{HodgesLong}]\label{thm:ryll}
A countable relational structure $\A$ is $\omega$-categorical
if and only if the automorphism group of $\A$ has for each $k$ finitely many orbits of $k$-tuples.
\end{thm}
\noindent It is a well-known consequence of the proof of Theorem~\ref{thm:ryll}, as presented for instance in~\cite{HodgesLong}, that $\A$ is $\omega$-categorical if and only if Inv$(\text{Aut}(\A)) = \langle \A\rangle_{\mathrm{fo}}$ (see, e.g., \cite{Bodirsky}). One direction persists in the realm of the primitive positive, as attested to by the following.
\begin{thm}[see \cite{Geiger,BoKaKoRo,BodirskyNesetrilJLC}] 
\label{thm:inv-pol-finite-omegacat}
When $\A$ is finite or $\omega$-categorical, $\Inv(\Pol(\A))= \langle \A\rangle_{\mathrm{pp}}$.
\end{thm}
\noindent This characterization is not tight, i.e., there are infinite non-$\omega$-categorical strutures $\A$ for which $\Inv(\Pol(\A))= \langle \A\rangle_{\mathrm{pp}}$ \cite{RomovPosPrimStruc}. 

\subsection{The constraint satisfaction problem}

For a relational structure $\A$ over a finite signature, $\mathrm{CSP}(\A)$ is the computational problem to decide whether a given pp-sentence is true in $\A$. Proof of the following is straightforward (see, e.g., \cite{JeavonsClosure}).
\begin{prop}
For $\A$ and $\A'$ with the same domain, such that $\langle \A \rangle_{\mathrm{pp}} \subseteq \langle \A' \rangle_{\mathrm{pp}}$, we have that $\mathrm{CSP}(\A)$ is polynomial time many-to-one reducible to $\mathrm{CSP}(\A')$.
\end{prop}
We remark that this result holds for logspace reductions, though this is harder to see and requires the celebrated result of \cite{Reingold}. In light of this observation, together with Theorem~\ref{thm:inv-pol-finite-omegacat}, we may use the sets $\Pol(\A)$ to classify the computational complexity of $\mathrm{CSP}(\A)$, and a most successful research program has run in this direction (see \cite{JeavonsClosure,Conservative,Bulatov}, and \cite{CSPSurveys} for a survey).

Sets of the form $\Pol(\A)$ are always \emph{clones} (for definitions, see \cite{Szendrei}), and the machinery of Clone Theory can be brought to bear on the classification program for CSPs (e.g., the tame congruence theory from \cite{HobbyMcKenzie} as laid out in \cite{DBLP:conf/dagstuhl/BulatovV08}). It often transpires that instances of the CSP with low complexity can be explained by the presence of certain polymorphisms on the template. When $\A$ is finite, the class of problems $\mathrm{CSP}(\A)$ is conjectured to display complexity dichotomy between those problems that are in P and those that are NP-complete (a remarkable property given the breadth of CSP problems together with the result of Ladner that NP itself does not possess the dichotomy, so long as P$\neq$NP \cite{Ladner}). While the \emph{dichotomy conjecture} was formulated independently of the algebraic method \cite{FederVardi}, a conjecture as to exactly where the boundary sits was given in the algebraic language \cite{JBK}. 

In the case where $\A$ is infinite but $\omega$-categorical, the connection of Theorem~\ref{thm:inv-pol-finite-omegacat} has been used to good effect in the complexity classification of, e.g., temporal CSPs in \cite{tcsps}. In that case a dichotomy between P and NP-complete was again observed. For $\omega$-categorical templates in general, it is known that there are structures whose CSP is undecidable \cite{BodirskyGrohe} and of various complexities \cite{BodirskyGrohe} (even coNP-complete). While the algebraic machinery has proved very powerful in the finite and $\omega$-categorical cases, for infinite templates that are not $\omega$-categorical, no such technology has thus far been developed.

\section{Primitive positive definability of formulas}
To show hardness of CSP$(\A)$, we often try to prove that there is a finite signature reduct $\A'$ of $\langle \A\rangle_{\mathrm{pp}}$ such that CSP$(\A')$ is NP-hard.
An important set of relations that contains the set of all pp-definable relations $\langle \A\rangle_{\mathrm{pp}}$ is the set of all fo-definable relations $\langle \A\rangle_{\mathrm{fo}}$. For every structure $\A$ of cardinality greater than one there are fo-definable relations yielding an NP-hard CSP,
and these relations are usually good candidates for proving hardness (e.g. $R:=\{0,1\}^3 \setminus \{(0,0,0),(1,1,1)\}$ on boolean domains; $\{(a_1,a_2):a_1 \neq a_2\}$ on finite domains $A$, $|A|\geq 3$; $\{(a_1,a_2,a_3):(a_1=a_2 \wedge a_2\neq a_3)\vee (a_1\neq a_2 \wedge a_2= a_3)\}$ on infinite domains $A$). Therefore, it is natural and important to understand which fo-definable relations are pp-definable in $\A$. 

\subsection{Equivalent templates}

In this section we show that, for every problem CSP$(\A)$, we can find a relational structure $\M$ for which CSP($\A$) = CSP($\M$) where infinitary polymorphisms exactly characterize pp-definability of fo-definable relations. We will do this by building a model of $\mathrm{Th}(\A)$ that is highly saturated. 
\begin{defi}
A $\tau$-structure $\M$ has the \emph{homomorphism lifting property} if, for any finite $k$, $a_1,\ldots,$ $a_k \in \M^\omega$ and $b_1,\ldots,b_k \in \M$ such that all pp-($\tau \cup \{c_1,\ldots,c_k\}$)-sentences true in $(\M^\omega;a_1,\ldots,a_k)$ are true in $(\M;b_1,\ldots,b_k)$, then there is a homomorphism $f:(\M^\omega;a_1,\ldots,a_k) \rightarrow $$ (\M;b_1,\ldots,b_k)$.
\end{defi}
\noindent The most natural of structures with the homomorphism lifting property are those that are of large cardinality and saturated.
\begin{lem}
\label{lem:sat-has-hlp}
If $\M$ is a saturated structure of cardinality $\kappa =  \kappa^\omega$, then $\M$ has the homomorphism lifting property.
\end{lem}
\proof
This follows immediately from Lemma~\ref{lem:sat-hom}, since $|M^\omega| \leq |M|$.
\qed
\noindent We remark that the continuum has the property of Lemma~\ref{lem:sat-has-hlp} -- that is $2^\omega = (2^\omega)^\omega$. On the assumption of the continuum hypothesis, we could simply work with large saturated structures; because then, given an infinite model $\A$, we can assume the existence of an elementary extension $\M$ that is of cardinality $2^\omega$ and saturated \cite{Hodges}. Without such a set-theoretic assumption, we can still construct a rather unwieldy model as follows.
\begin{lem}
\label{lem:monster-model}
For every $\tau$-structure $\A$ there is an elementary extension $\M \succeq \A$ that is $\omega$-saturated and has the homomorphism lifting property.
\end{lem}
\proof
We will build $\M$ by transfinite induction, as the union of a chain of length $\aleph_1$. Set $\M_0:=\A$. For successor ordinals $\gamma+1$, we take an elementary extension $\M_{\gamma+1} \succeq \M_{\gamma}$ that is $|M_\gamma|^\omega$-saturated (such always exists, see Corollary 8.2.2 \cite{Hodges}). For limit ordinals $\lambda$, set $\M_{\lambda}:=\bigcup_{\alpha<\lambda} \M_{\alpha}$; finally, let $\M:=\M_{\aleph_1}$.

$\M$ is $\omega$-saturated by construction. It remains to prove that $\M$ has the homomorphism lifting property. Consider the $b_1,\ldots,b_k \in M$ and $a_1,\ldots,a_k \in M^\omega$. The set of coordinates (of $M$) involved here,
\[
\begin{array}{r}
A:=\{b_1,\ldots,b_k,a_1(1),a_1(2),\ldots, a_2(1), a_2(2),\ldots,\ldots, a_k(1),a_k(2),\ldots \},
\end{array}
\]
is of size $\leq \omega$. It follows that there is some $\mu< \aleph_1$ such that $A \subseteq M_\mu$ (this is why the chain used in the construction of $\M$ is of length $\aleph_1$).

Suppose we are given $a_1,\ldots,a_k \in \M^\omega$ and $b_1,\ldots,b_k \in \M$ such that all pp-($\tau \cup \{c_1,\ldots,c_k\}$)-sentences true in $(\M^\omega;a_1,\ldots,a_k)$ are true in $(\M;b_1,\ldots,b_k)$.
Let $f_{-1}$ be the partial map from $M^\omega$ to $M$ sending $a_1,\ldots,a_k$ to $b_1,\ldots,b_k$ (well-defined as all equalities are pp). We first argue that all pp-($\tau \cup \{c_1,\ldots,c_k\}$)-sentences true in $({\M_\mu}^\omega;a_1,\ldots,a_k)$ are true in $(\M_{\mu+1};f_{-1}(a_1),\ldots,f_{-1}(a_k))$. Let $\phi$ be such a sentence. Then:
\[
\begin{array}{rcl}
& ({\M_\mu}^\omega;a_1,$ $\ldots,a_k) \models \phi & \Rightarrow \\
\mbox{for each $i$ } & ({\M_\mu};a_1(i),\ldots,a_k(i)) \models \phi & \Rightarrow^{a}  \\
\mbox{for each $i$ } & ({\M};a_1(i),\ldots,a_k(i)) \models \phi & \Rightarrow \\
& (\M^\omega;a_1,$ $\ldots,a_k) \models \phi & \Rightarrow^{b} \\
& (\M;f_{-1}(a_1),\ldots,f_{-1}(a_k)) \models \phi & \Rightarrow^{c}  \\
& (\M_{\mu+1};f_{-1}(a_1),\ldots,f_{-1}(a_k)) \models \phi. & \\
\end{array}
\]
$(a)$ since $\M_{\mu} \preceq \M$; $(b)$ by hypothesis; $(c)$ since $\M \succeq \M_{\mu+1}$.
\noindent It follows from Lemma~\ref{lem:sat-hom} that there is a homomorphism $f_0:{\M_\mu}^\omega \rightarrow \M_{\mu+1}$ extending $f_{-1}$.

We will now proceed with a transfinite induction up to $\aleph_1$. For successor ordinals, $\gamma+1$, suppose that we have a homomorphism $f_\gamma:{\M_{\mu+\gamma}}^\omega \rightarrow \M_{\mu+\gamma+1}$. We will build a homomorphism $f_{\gamma+1}:{\M_{\mu+\gamma+1}}^\omega \rightarrow \M_{\mu+\gamma+2}$ extending $f_\gamma$. 
Since $f_\gamma$ is a homomorphism, all pp-($\tau \cup \{c_\alpha : \alpha < |M_{\mu+\gamma}|^\omega\}$)-sentences true in $({\M_{\mu+\gamma}}^\omega;\langle a_\alpha \rangle_{\alpha <|M_{\mu+\gamma}|^\omega})$ 
-- where $\langle a_\alpha \rangle_{\alpha <|M_{\mu+\gamma}|^\omega}$ 
well-orders the elements of ${\M_{\mu+\gamma}}^\omega$ -- 
are true in $({\M_{\mu+\gamma+1}};\langle f_\gamma(a_\alpha) \rangle_{\alpha <|M_{\mu+\gamma}|^\omega})$.
It follows that all pp-($\tau \cup \{c_\alpha : \alpha < |M_{\mu+\gamma}|^\omega\}$)-sentences true in $({\M_{\mu+\gamma+1}}^\omega;\langle a_\alpha \rangle_{\alpha <|M_{\mu+\gamma}|^\omega})$ are true in $({\M_{\mu+\gamma+2}};\langle f(a_\alpha) \rangle_{\alpha <|M_{\mu+\gamma}|^\omega})$ -- let $\phi$ be such a sentence, we give the derivation again:
\[
\begin{array}{rcl}
& ({\M_{\mu+\gamma+1}}^\omega;\langle a_\alpha \rangle_{\alpha <|M_{\mu+\gamma}|^\omega}) \models \phi & \Rightarrow \\
\mbox{for each $i$ } & ({\M_{\mu+\gamma+1}};\langle a_\alpha(i) \rangle_{\alpha <|M_{\mu+\gamma}|^\omega}) \models \phi  & \Rightarrow \\
\mbox{for each $i$ } & ({\M_{\mu+\gamma}};\langle a_\alpha(i) \rangle_{\alpha <|M_{\mu+\gamma}|^\omega}) \models \phi & \Rightarrow \\
& ({\M_{\mu+\gamma}}^\omega;\langle a_\alpha \rangle_{\alpha <|M_{\mu+\gamma}|^\omega}) \models \phi & \Rightarrow \\
& ({\M_{\mu+\gamma+1}};\langle f(a_\alpha) \rangle_{\alpha <|M_{\mu+\gamma}|^\omega}) \models \phi & \Rightarrow \\
& ({\M_{\mu+\gamma+2}};\langle f(a_\alpha) \rangle_{\alpha <|M_{\mu+\gamma}|^\omega}) \models \phi . \\
\end{array}
\]
Now we can use Lemma~\ref{lem:sat-hom} to derive some homomorphism $f_{\gamma+1}:{\M_{\mu+\gamma+1}}^\omega \rightarrow \M_{\mu+\gamma+2}$ extending $f_\gamma$. For limit ordinals $\lambda$, set $f_\lambda:=\bigcup_{\alpha<\lambda} f_\alpha$.

Finally, we arrive at the homomorphism $f_{\aleph_1}:\M^\omega \rightarrow \M$, which has the desired property.
\qed
We describe the model built as in the previous lemma as a \emph{monster} model (this term is used with different connotation in set theory and model theory).  

Let $\langle \A\rangle_{\mathrm{pp}\infty}$ be the set of relations pp-definable on $\A$, \emph{possibly involving infinitary conjunction} (of pp-formulas in a finite number of free variables). Because we will use it again later, we give the following lemma in its strongest form.
\begin{lem}
\label{lem:pp+}
For all structures $\A$, $\langle \A \rangle_{\mathrm{pp \infty}} \subseteq \Inv(\Pol^\infty(\A))$.
\end{lem}
\proof
We argue by induction on the term-complexity of the formula. Let $f: A^\alpha \rightarrow A$ be a polymorphism of $\A$.

(Base Case.) $\phi(\overline{v}):=R(\overline{v})$. Trivial.

(Inductive Step.) There are two subcases. In the following, suppose $\overline{v}$ is an $m$-tuple. Let $\langle \overline{a}_\beta \rangle_{\beta < \alpha}$, be a sequence of $m$-tuples from $\A$ such that $\A \models \phi(\overline{a}_\beta)$, for all $\beta$.

(Existential Quantification.) $\phi(\overline{v}):= \exists u. \psi(\overline{v},u)$. Suppose we have $\A \models \phi(\overline{a}_\beta)$ for each $\beta < \alpha$. From each $\A \models \exists u. \psi(\overline{a}_\beta,u)$, derive the witness $a'_\beta$ for $u$ and use the inductive hypothesis to deduce that 
$\A \models \psi(f(\langle \overline{a}_\beta \rangle_{\beta < \alpha}), f(\langle a'_\beta \rangle_{\beta< \alpha}))$.
It follows that $\A \models \exists u. \psi(f(\langle \overline{a}_\beta \rangle_{\beta < \alpha}),u)$ and we are able to deduce $\A \models \phi(f(\langle \overline{a}_\beta \rangle_{\beta < \alpha}))$.

(Infinite Conjunctions.) $\phi(\overline{v}):= \bigwedge_{\mu < \gamma} \psi_\mu(\overline{v})$. Suppose we have $\A \models \phi(\overline{a}_\beta)$ for each $\beta < \alpha$.  Then for each $\mu < \gamma$ and $\beta < \alpha$ we have $\A \models \psi_\mu(\overline{a}_\beta)$. By inductive hypothesis, we have each $\A \models \psi_\mu(f(\langle \overline{a}_\beta \rangle_{\beta <\alpha}))$. The result $\A \models \phi(f(\langle \overline{a}_\beta \rangle_{\beta <\alpha}))$ follows.
\qed
We are now ready for the main result of this section.
\begin{thm}\label{thm:inv-pol-inf}
Let $\A$ be a structure, over a countable\footnote{
The restriction to countable signatures allows us to consider polymorphisms of at most arity $\omega$. Were we to consider signatures of infinite cardinality $\alpha$ then the theorem would hold with polymorphisms of arity $\alpha$ (under a slightly different definition of the homomorphism lifting property involving $\alpha$ in the exponent).
}
signature, that is $\omega$-saturated and has the homomorphism lifting property.
Then a fo-definable relation $R$ is preserved by the $\omega$-polymorphisms of
$\A$ if and only if $R$ is pp-definable in $\A$, i.e.
\[ \Inv(\Pol^\omega(\A))  \cap \langle \A \rangle_{\mathrm{fo}} = \langle \A \rangle_{\mathrm{pp}}.\] 
\end{thm}
\proof
(Backwards.)
That pp-formulas are preserved by $\omega$-polymorphisms in any structure is a special case of Lemma~\ref{lem:pp+}.

(Forwards.) Suppose that $R$ is a $k$-ary relation that is preserved by all $\omega$-polymorphisms of $\A$ and that has a first-order definition $\phi$ in $\A$. Let 
\[
\begin{array}{r}
\Psi:= \{ \psi(x_1,\ldots,x_k): \mbox{ $\psi$ is a pp-$\tau$-formula s.t. } \A \models \phi(x_1,\ldots,x_k) \rightarrow \psi(x_1,\ldots,x_k) \}.
\end{array}
\]
We first show, for all $b_1,\ldots,b_k \in A$, that, if $\A \models \Psi(b_1,\ldots,b_k)$, then $\A \models \phi(b_1,\ldots,b_k)$.

Take $b_1,\dots,b_k \in A$ \mbox{s.t.}
$\A \models \psi(b_1,\dots,b_k)$ for each $\psi \in \Psi$; if such elements do
not exist there is nothing to show.
Let $U$ be the set of all pp-$\tau$-formulas $\theta(x_1,\dots,x_k)$ 
such that $\A \models \neg \theta(b_1,\dots,b_k)$. $U$ contains $\bot$, $\bot \wedge \bot$ etc., and so we may assume $U$ to be countably infinite. Let $(\theta_i)_{i < \omega}$ be an enumeration of $U$.
We claim that for every $\theta_i \in U$
there exists a $k$-tuple $\overline{a}^{\theta_i}:=(a^{\theta_i}_1,\ldots,a^{\theta_i}_k)$ from $A$
such that $\A \models \neg \theta_i(a^{\theta_i}_1,\dots,a^{\theta_i}_k) \wedge \phi(a^{\theta_i}_1,\dots,a^{\theta_i}_k)$. 
Otherwise, $\A \models \phi(x_1,\dots,x_k) \rightarrow \theta_i(x_1,\ldots,x_k)$, and we derive $\theta_i \in \Psi$ and the consequent contradiction $\A \models \theta_i(b_1,\ldots,b_k)$.

Consider the $k$-tuple $\overline{a}:= \prod_{i <\omega} \overline{a}^{\theta_i}$ in $\A^\omega$.
Observe that every pp-$\tau$-formula $\chi(x_1,\dots,x_k)$ s.t. $\A^\omega \models \chi(\overline{a})$ is s.t. $\A \models \chi(b_1,\dots,b_k)$. 
To see this, suppose that $\A \models \neg \chi(b_1,\ldots,b_k)$. Therefore $\chi \in U$, 
and by choice of $\overline{a}^\chi$ we have $\A \models \neg \chi(\overline{a}^\chi)$.
But then $\A^\omega \models \neg \chi(\overline{a})$.

Now, we have just shown that all pp-($\tau \cup \{c_1,\ldots,c_k\}$)-sentences that hold on $(\A^\omega;\overline{a})$ also hold on $(\A;b_1,\ldots,b_k)$. Since $\A$ has the homomorphism lifting property, the existence of a homomorphism $f:(\A^\omega;\overline{a}) \rightarrow (\A;b_1,\ldots,b_k)$ follows from our definitions. But $f$ is an $\omega$-polymorphism of $\A$, which preserves $\phi$, and hence we derive $\A \models \phi(b_1,\ldots,b_k)$.

It remains to be shown that $\Psi$ is equivalent on $\A$ to a single
pp-formula. Note that $\Psi(c_1,\dots,c_k)$ $\cup \{\neg \phi(c_1,\dots,c_k)\} \cup \mathrm{Th}(\A)$ is unsatisfiable; for otherwise there is a $\B \models \mathrm{Th}(\A)$ and $b'_1,\ldots,b'_k \in B$, \mbox{s.t.} $(\B;b'_1,\ldots,b'_k) \models \Psi(c_1,\dots,c_k)$ and $(\B;b'_1,\ldots,b'_k) \models \neg \phi(c_1,\dots,c_k)$. Since $\A$ is $\omega$-saturated, this yields some $b''_1,\ldots,b''_k \in A$ such that both $(\A;b''_1,\ldots,b''_k)$ $\models$ $\Psi(c_1,\dots,c_k)$ and $(\A;b''_1,\ldots,b''_k) \models \neg \phi(c_1,\dots,c_k)$, which is a contradiction.
By compactness of first-order logic 
there is  a finite subset $\Psi'$ of $\Psi$
such that $\Psi'(c_1,\dots,c_k) \cup \{\neg \phi(c_1,\dots,c_k)\} \cup \mathrm{Th}(\A)$ is unsatisfiable, \mbox{i.e.} $\Psi'(c_1,\ldots,c_k) \cup \mathrm{Th}(\A) \models \phi(c_1,\ldots,c_k)$. Set $\psi'(x_1,\ldots,x_k):= \bigwedge_{\psi \in \Psi'} \psi(x_1,\ldots,x_k)$, to derive $\mathrm{Th}(\A) \models \psi'(x_1,\ldots,x_k) \rightarrow \phi(x_1,\ldots,x_k)$. Since $\A \models \phi(x_1,\ldots,x_k) \rightarrow \psi'(x_1,\ldots,x_k)$ by construction, the result follows.
\qed
\begin{cor}
Let $\A$ be any structure with finite relational signature. Then there exists a structure $\M$ such that CSP($\A$) = CSP($\M$), and such that an fo-definable relation $R$ is pp-definable in $\M$ if and only if $R$ is preserved by all $\omega$-polymorphisms of $\M$.
\end{cor}
\proof
By Lemma~\ref{lem:monster-model}, there is a monster elementary extension of $\M \succeq \A$ with the homomorphism lifting property. We now apply Theorem~\ref{thm:inv-pol-inf}.
\qed
\noindent In the parlance of \cite{Poizat:AlgebresDePost}, the following may be seen as the ``global'' analog of Theorem~\ref{thm:inv-pol-inf}.
\begin{cor}
An fo-formula $\phi$ is preserved by the $\omega$-polymorphisms of all elementary extensions of $\A$ if and only if $\phi$ is pp-definable in $\A$.
\end{cor}
\proof
(Backwards.) Follows from Lemma~\ref{lem:pp+}.

(Forwards.) Since $\phi$ is preserved by the $\omega$-polymorphisms of the monster elementary extension $\M \succeq \A$ constructed in Lemma~\ref{lem:monster-model}, it follows from Theorem~\ref{thm:inv-pol-inf} that $\phi$ is pp-definable on $\M$. But this is a fortiori a pp-definition on $\A$.
\qed
\begin{cor}
Let $T$ be an uncountably categorical fo-theory, and $\A$ a model of $T$ of cardinality $\kappa=\kappa^\omega$.
Then $\Inv(\Pol^\omega(\A))  \cap \langle \A \rangle_{\mathrm{fo}} = \langle \A \rangle_{\mathrm{pp}}$.
\end{cor}
\proof
It is well-known that uncountable models of uncountably categorical theories are saturated in their own
cardinality (Fact 1.2. in \cite{Zilber}).
Hence, the statement follows from Lemma~\ref{lem:sat-has-hlp} and Theorem~\ref{thm:inv-pol-inf}.
\qed

\subsection{Tightness of Theorem~\ref{thm:inv-pol-inf}}
One might be interested in the following potential strengthenings
of Theorem~\ref{thm:inv-pol-inf}.
\begin{enumerate}[(1)]
\item To derive the statement for arbitrary relations (not just for fo-definable relations).
\item To assume preservation under finitary polymorphisms (not infinitary polymorphisms).
\item To show the statement for arbitrary models of $\mathrm{Th}(\A)$ (not just for structures with the homomorphism lifting property).
\end{enumerate}
\noindent The following proposition shows that each of these assumptions is necessary.
\begin{prop} \
\label{prop:three-parts}
\begin{enumerate}[\em(1)]
\item There is a saturated structure $\A_{\mathrm{sat}}$ of cardinality $2^\omega$ such that $\Inv(\Pol^\omega(\A_{\mathrm{sat}})) \neq \langle \A_{\mathrm{sat}} \rangle_{\mathrm{pp}}$.
\item There is a saturated structure $\A_{\mathrm{sat}}$ of cardinality $2^\omega$ such that $\Inv(\Pol(\A_{\mathrm{sat}}))  \cap \langle \A_{\mathrm{sat}} \rangle_{\mathrm{fo}} \neq \langle \A_{\mathrm{sat}} \rangle_{\mathrm{pp}}$.
\item There is a structure $\A$ such that $\Inv(\Pol^\omega(\A))  \cap \langle \A \rangle_{\mathrm{fo}} \neq \langle \A \rangle_{\mathrm{pp}}$.
\end{enumerate}
\end{prop}
\proof

\textbf{Necessity of intersection with fo}.
Let us consider the model $\A=:(\mathbb{Q};+,1,(u=v \vee x=y))$. By Lemma~\ref{lem:essentially-unary}, the infinitary polymorphisms of this structure are equivalent to its endomorphisms, and, in the presence of a fixed $1$, it can easily be seen that its only endomorphism is the identity (indeed, there is a pp-definition of each of the rationals from $1$ and $+$). It follows that all subsets of $\mathbb{Q}$ are in $\Inv(\Pol^\omega(\A))$, yet $\langle \A\rangle_{\textrm{pp}}$ must be countable. Of course, $\A$ is neither saturated nor of cardinality $2^\omega$. But the continuum of subsets of $\mathbb{Q}$ will remain $\Inv\mbox{-}\Pol^\omega$ in a saturated model of $\mathrm{Th}(\A)$ of such cardinality (a copy of $\A$ sits elementary in all models of its theory). The existence of a saturated model of $\mathrm{Th}(\A)$ of cardinality $2^\omega$ follows from this theory's strong minimality (Fact 1.2. in \cite{Zilber}).

\textbf{Necessity of infinitary polymorphisms}.
Let $\{U_i : i \in \omega\}$ be a set of unary relations. Consider the model $\A:=( \mathbb{N}; U_i : i \in \omega )$, involving a countable set of unary relations, defined by $U_i:=\mathbb{N} \setminus \{0,i\}$. Diagrammatically, 

\begin{center}
\begin{tabular}{|c|c|c|c|c|} \hline
   & $U_1$        & $U_2$         & $U_3$        & $\cdots$ \\ \hline
$0$ & $\times$ & $\times$ & $\times$ &  $\cdots$ \\ \hline
$1$ & $\times$ & $\surd$   & $\surd$   &  $\cdots$ \\ \hline
$2$ & $\surd$   & $\times$ & $\surd$   &  $\cdots$ \\ \hline
$3$ & $\surd$   & $\surd$   & $\times$ &  $\cdots$ \\ \hline
\vdots & \vdots & \vdots & \vdots & \\ \hline 
\end{tabular}
\end{center}

\noindent Consider the fo-definable unary relation $P(v):=U_1(v) \vee U_2(v)$, i.e. $P:=\mathbb{N}\setminus \{0\}$. It is straightforward to verify that $P$ is closed under the finitary polymorphisms of $\A$ and is not pp-definable over $\A$. Note that $P$ is not preserved under the infinitary polymorphism $f:\mathbb{N}^\omega \rightarrow \mathbb{N}$ of $\A$ defined by $f(\overline{w})=0$, if $\overline{w}$ contains all elements of $\mathbb{N} \setminus \{0\}$, and $f(\overline{w})=w_0$ (the first element of the sequence $\overline{w}$), otherwise.
Again, these properties will remain if we move to a saturated model $\A_{\mathrm{sat}}$ of cardinality $2^\omega$ (such a model will simply be $\A$ augmented with a continuum of elements for which all of the relations $\{U_i : i \in \omega\}$ hold).

We now detail a finite signature variant of the above structure that also serves as a suitable (counter)example. Consider the signature $\langle E,R \rangle$ involving two binary relations, edge and red edge. Let the structure $\A$ contain
\begin{iteMize}{$\bullet$}
\item a directed $\omega$-$E$-path: i.e., vertices $\{(0,i) : i < \omega\}$ and $E$-edges $\{((0,i),(0,i+1)) : i < \omega\}$,
\item[] and for each $j<\omega$: 
\item a directed $\omega$-$E$-path with overlaid undirected $R$-path omitting only the $j$th edge: i.e. vertices $\{(j,i) : i < \omega \}$ with $E$-edges $\{((j,i),(j,i+1)) : i < \omega\}$ and $R$-edges $\{((j,i),(j,i+1)), ((j,i+1),(j,i))  : i < \omega, i+1 \neq j\}$.
\end{iteMize}
\noindent Consider the first-order definable unary relation $P(v):=\exists x,y. R(v,x) \vee (E(v,x) \wedge R(x,y))$. It is not hard to verify that $P$ is preserved by the finitary polymorphisms of $\A$, but is not pp-definable over $\A$ (as it is not preserved by the $\omega$-polymorphisms of $\A$). The saturated elementary extension of $\A$ of cardinality $2^\omega$ contains $2^\omega$ copies of each one of the countably infinite set of paths just described. It remains easy to see that $P$ is preserved by $\A_{\mathrm{sat}}$'s finitary but not $\omega$-ary polymorphisms. 

\textbf{Necessity of highly saturated structures}.
Consider the structure $\mathfrak{A}:=(\mathbb{Q}; x=1, x<0, S_2(x,y) )$, where $S_2:=\{ (x,y) : 2x<y, 0<y\leq 1 \}$. Now, $x \leq 0$ is clearly fo-definable in $\mathfrak{A}$ (this is from $x<0$; actually $x \leq 0$ is fo-definable from just $S_2$, so $x<0$ is not needed as an extensional relation). It is also in $\Inv(\Pol^\omega(\mathfrak{A}))$, being definable by the following infinite conjunction of pp-formulas in one free variable (see Lemma~\ref{lem:pp+}).
\[ \bigwedge_{i \in \omega} \exists z \ \exists y_1,\ldots,y_i. \ S_2(x,y_1) \wedge S_2(y_1,y_2) \wedge \ldots \wedge S_2(y_i,z) \wedge z=1. \]
We will now argue that it is not pp-definable.

\noindent \textbf{Lemma.}
Let $\overline{x}:=(x_1,\ldots,x_k)$ and suppose that $\phi(\overline{x}) \in \langle \mathfrak{A} \rangle_{\mathrm{pp}}$. If $\mathfrak{A} \models \phi(\overline{a})$ and $a_{\lambda_1},\ldots,a_{\lambda_j}$ list exactly the elements of $\overline{a}$ that are $0$, then there exists $\epsilon>0$ such that, for all $\epsilon \geq \delta \geq 0$, $\mathfrak{A} \models \phi(\overline{a}[a_{\lambda_1}/\delta,\ldots,a_{\lambda_j}/\delta])$.
\proof
By induction on the term complexity of $\phi$.

(Base Cases.) $\phi$ is an atom. The statement is trivially true if $\phi(x):=$ $x=1$, $x<0$ or $x=x$. Suppose $\phi(x_1,x_2):=S_2(x_1,x_2)$; if $S_2(a_1,a_2)$, then only $a_1$ could be zero. Set $\epsilon:=a_2/2$.

(Inductive Step.) There are two subcases.

$\phi(\overline{x}):=\psi_1(\overline{x}) \wedge \psi_2(\overline{x})$. There exist respective witnesses $\epsilon_1$ and $\epsilon_2$ for $\psi_1(\overline{a})$ and $\psi_2(\overline{a})$: we may set $\epsilon:=\min\{\epsilon_1,\epsilon_2\}$ as the witness for $\phi(\overline{a})$.

$\phi(\overline{x}):=\exists y. \psi(y,\overline{x})$. If $\A \models \phi(\overline{a})$ holds, then we may choose a $b$ \mbox{s.t.} $\A \models \psi(b,\overline{a})$. By inductive hypothesis, there exists an appropriate $\epsilon$ for $\psi(b,\overline{a})$ and this may also be used for $\phi(\overline{a})$.
\qed
\noindent That $x \leq 0$ is not pp-definable is a trivial consequence of the lemma, for suppose it were defined by $\phi(x)$. Since $\phi(0)$ holds, we may derive the contradiction that $\phi(\epsilon)$ holds for some $\epsilon >0$. Note that the first part of the inductive step in the previous lemma would fail for infinite conjunctions. Finally, suppose $\mathfrak{A}_{\mathrm{sat}}$ were a saturated model of $\mathrm{Th}(\mathfrak{A})$ of cardinality $2^\omega$. While we have $\langle \mathfrak{A} \rangle_{\mathrm{fo}} \cap \Inv(\Pol^\omega(\mathfrak{A})) \neq \langle \mathfrak{A} \rangle_{\mathrm{pp}}$, we must have $\langle \mathfrak{A}_{\mathrm{sat}} \rangle_{\mathrm{fo}} \cap \Inv(\Pol^\omega(\mathfrak{A}_{\mathrm{sat}})) = \langle \mathfrak{A}_{\mathrm{sat}} \rangle_{\mathrm{pp}}$. We note that $x \leq 0$ is not $\Inv(\Pol^\omega(\mathfrak{A}_{\mathrm{sat}}))$.
\qed

\section{Existential-positively closed models}
\label{sec-epc1}

In this section, we define existential-positively closed models and demonstrate their applications in constraint satisfaction.

\subsection{Definitions and basic results}
\label{sec-epc2}

We begin by stating  some basic facts
regarding existential-positively closed models. They
are the positive analogs of existentially closed models (the latter are treated in great detail in Section 7 of \cite{Hodges}, which is Section 8 of \cite{HodgesLong}),
and have been studied under the name of existentially closed
models in a recent paper on positive model theory by Ben-Yaacov~\cite{BenYaacov}.

\begin{defi}
A model $\mathfrak A$ is 
\emph{existential-positively closed for $T$} -- or short \emph{epc} -- 
iff $\mathfrak{A} \models T$ and for any homomorphism $h$ from $\mathfrak A$ into another model
$\mathfrak B$ of $T$, any tuple $\bar a$ from $A$,
and any existential positive formula $\phi$ with ${\mathfrak B} \models \phi(h(\bar a))$ we have that ${\mathfrak A} \models \phi(\bar a)$.
\end{defi}
\noindent Note that we could equivalently have used primitive positive formulas in the previous definition. To show the existence of certain epc models we apply
the direct limit construction.

\begin{defi}
\label{def:limit}
Let $\tau$ be a relational signature, 
and 
let $(\bA_i)_{i< \kappa}$ be a sequence of $\tau$-structures of cardinality $\leq \kappa$ such that there are homomorphisms $f_{ij}: \bA_i \rightarrow \bA_j$
with $f_{jk} \circ f_{ij} = f_{ik}$ for every $i \leq j \leq k$.
Then the \emph{direct limit} $\lim_{i<\kappa} \bA_i$ (with respect to the homomorphisms $f_{ij}$) is the
$\tau$-structure $\bA$ defined as follows. 
The domain $A$ of $\bA$
comprises the equivalence classes of the equivalence relation
$\sim$ defined on $\bigcup_{i < \kappa} A_i$ 
by setting
$x_i \sim x_j$ for $x_i \in A_i$ and $x_j \in A_j$ iff there is a $k$ such that
$f_{ik}(x_i) = f_{jk}(x_j)$.
Let $g_i: A_i \rightarrow A$ be the \emph{limit homomorphism}, \mbox{i.e.} the function that maps
$a \in A_i$ to the equivalence class of $a$ in $A$.
For $R \in \tau$, define $\bA \models R(\bar a)$ iff there
is a $k$ and $\bar b \in A_k$ such that $\bA_k \models R(\bar b)$
and $\bar a = g_k(\bar b)$. Note that limit homomorphisms behave well with respect to their homomorphism family (\mbox{i.e.} $g_i = g_j \circ f_{ij}$).
\end{defi}
\noindent The direct limits defined above can be seen as a positive variant 
of the basic model-theoretic notion of a \emph{union of chains} (see Section 2.4 in~\cite{HodgesLong}); we essentially replace embeddings in chains by homomorphisms. Unions of chains preserve 
$\forall \exists$-sentences; the analogous
statement for direct limits is as follows.
A sentence is called \emph{$\forall\exists^+$} if
it is a universally quantified positive boolean combination
of existential positive formulas and negated atomic formulas.
We could have equivalently defined $\forall\exists^+$-formulas as conjunctions of 
universally quantified 
disjunctions
of primitive positive formulas and negated atomic formulas.
It is easy to see that every $\forall\exists^+$-formula can be re-written into such a formula.
\begin{prop}[see Theorem 2.4.6 in~\cite{HodgesLong}]
\label{prop:direct-product-preservation}
Let $\A$ be the direct limit of $(\bA_i)_{i< \kappa}$; 
if $\phi$
is $\forall\exists^+$
such that $\A_i \models \phi$ for all $i$,
then $\A \models \phi$.
\end{prop}
\begin{lem}\label{lem:epc-limits}
The class of all epc models for a theory $T$ is closed under
direct limits.
\end{lem}
\proof
Suppose that $\bA = \lim_{\alpha<\kappa} \bA_\alpha$ for a sequence $(\bA_\alpha)_{\alpha < \kappa}$ of models that are epc for $T$, $\bar a$ is a tuple from $\bA$, $\phi$ an ep-formula, and $h$ is a homomorphism from $\bA$ into another model of $T$ such that $\bB \models \phi(h(\bar a))$.
Then there exists a $\alpha < \kappa$ such that $\bar a=g_\alpha(\bar a')$ for $\bar a'$ from $\bA_\alpha$ (where $g_\alpha$ is as in the definition of direct limits).
Note that $h \circ g_\alpha$ is a homomorphism from $\bA_\alpha$ to $\bB$, and
since $\bA_\alpha$ is a model epc for $T$,
$\bA_\alpha \models \phi(\bar a')$.
Since $g_\alpha$ preserves ep-formulas, we
thus also have that $\bA \models \phi(\bar a)$.
\qed
\begin{prop}[Essentially from~\cite{BenYaacov}]
\label{prop:existence-epc}
Let $\mathfrak A$ be a model of cardinality $\kappa$ of a set $T$ of $\forall\exists^+$ sentences. 
Then there is a homomorphism from $\mathfrak A$ to a  model $\mathfrak B$ that is epc for $T$ of cardinality $\leq \kappa$.
\end{prop}
\proof
Set ${\mathfrak B}_0: = \mathfrak A$. Let $i>0$ be a natural number and assume we have already constructed ${\mathfrak B}_{i-1}$ of cardinality $\leq\kappa$. Let 
$\{(\phi_\alpha,\bar a_\alpha) \; | \; \alpha < \kappa\}$ be a (not necessarily injective) enumeration of all pairs $(\phi,\bar a)$ where $\phi$ is existential positive with
free variables $x_1,\dots,x_n$, and $\bar a$ is an $n$-tuple
from $B_{i-1}$. 
We construct a sequence
$({\mathfrak B}_i^{\alpha})_{0 \leq \alpha < \kappa}$ of models of $T$ of cardinality $\leq\kappa$
and a coherent sequence
$(f_i^{\mu,\alpha})_{0 \leq \mu <\alpha < \kappa}$
of homomorphisms, where $f_i^{\mu,\alpha}:{\mathfrak B}_i^\mu\rightarrow{\mathfrak B}_i^\alpha$, as follows (coherent in the sense that, for $\alpha<\beta<\gamma$, $f_i^{\alpha,\gamma} = f_i^{\beta,\gamma} \circ f_i^{\alpha,\beta}$).

\smallskip

Set ${\mathfrak B}_i^0 = {\mathfrak B}_{i-1}$.
Now let $\alpha=\beta+1 < \kappa$ be a successor ordinal.
Let $\bar b_\beta$
be the image of $\bar a_\beta$ in ${\mathfrak B}_i^\beta$ under $f_i^{0,\beta}$ (in case $\beta=0$, set $\bar b_0:=\bar a_0$ ). If there is a model
$\bC$ of $T$ and a homomorphism $h: {\mathfrak B}_i^\beta \rightarrow \mathfrak C$ such that $\bC \models \phi_\beta(h(\bar b_\beta))$, then by the theorem of L\"owenheim-Skolem there is
also a model $\bC'$ of cardinality $\leq\kappa$ of $T$ and a
homomorphism $h': {\mathfrak B}_i^\beta \rightarrow \bC'$ such that $\bC' \models \phi_\beta(h'(\bar b_\beta))$.
Set ${\mathfrak B}^{\alpha}_i = \bC'$ and $f_i^{\mu,\alpha}= h'\circ f_i^{\mu,\beta} $.
Otherwise, if there is no such model $\bC$, we set ${\mathfrak B}^{\alpha}_i = {\mathfrak B}^{\beta}_i$ and $f_i^{\beta,\alpha} = \text{id}$ (the identity mapping) and $f_i^{\mu,\alpha} = f_i^{\mu,\beta}$.
Finally,  for limit ordinals $\alpha<\kappa$, set $\bB^\alpha_i =
\lim_{\mu<\alpha}\bB^\mu_i$
and let $f_i^{\mu,\alpha}$ be the corresponding limit homomorphism from $\bB^\mu_i$ to $\bB^\alpha_i$.

Let $\bB_i$ be
$\lim_{\alpha<\kappa} \bB_i^\alpha$ (with respect to $f_i^{\mu,\alpha}$) and let $g_{i-1}:\bB_{i-1}=\bB_i^0 \rightarrow \bB_i$ be the corresponding limit homomorphism.

Let $\bB = \lim_{i < \omega} \bB_i$. By construction, $\bB$ is of cardinality $\leq\kappa$, and it is a model of $T$ by Proposition~\ref{prop:direct-product-preservation}; let $h_i: \bB_i \rightarrow \bB$ 
for $i<\omega$ be the corresponding homomorphisms.

The structure $\bB$ is epc for $T$.
To verify this, let $g$ be a homomorphism from
$\bB$ to a model $\bC$ of $T$, and suppose that there is a tuple
$\bar b$ over $B$ and an existential positive formula $\phi$
such that $\bC \models \phi(g(\bar b))$. There is an $i < \omega$ and an $\bar a \in B_i$ such that
$h_i(\bar a)=\bar b$. Then $g \circ h_i$ is a homomorphism from $\bB_i$
to $\bC$, and by our construction we have that $\bB_{i+1} \models \phi(g_i(\bar a))$. Note that $h_{i+1}\circ g_i=h_i$. Thus, since $h_{i+1}$ preserves existential positive formulas,
we also have that $\bB \models \phi(\bar b)$, which is what
we had to show.
\qed
\begin{prop}\label{prop:epc-types}
Let $T$ be a theory, and let $\bA$ be a model of $T$.
Then $\bA$ is epc for $T$ if and only if every complete ep-$n$-type
of $\bA$ is a maximal ep-type of $T$.
\end{prop}
\proof
(Forwards.) Suppose $p(x_1,\ldots,x_n)$ is the complete ep-$n$-type, realized in $\bA$
by the tuple $(a_1,\ldots,a_n)$. Let $c_1,\dots,c_n$ be new constant symbols
that denote $a_1,\dots,a_n$ in $\bA$.
Let $\phi(x_1,\dots,x_n)$ be an existential positive formula such that
$T \cup p(c_1,\ldots,c_n) \cup \{\phi(c_1,\ldots,c_n)\}$ has a model $(\bC;b_1,$ $\ldots,b_n)$.
Now, let $(\bC_{\mathrm{sat}};b_1,\ldots,b_n)$
be an $|A|$-saturated model of $\mathrm{Th}(\bC;b_1,\ldots,b_n)$;
such a model always exists (see Corollary 8.2.2 \cite{Hodges}). Clearly $(\bC_{\mathrm{sat}};b_1,\ldots,b_n)$ is ep-$|A|$-saturated, and all existential positive formulas that are true on $(\bA;a_1,\ldots,a_n)$ are true on $(\bC_{\mathrm{sat}};b_1,\ldots,b_n)$. By Lemma~\ref{lem:sat-hom}, there is a homomorphism $h$ from $(\bA;a_1,\ldots,a_n)$ to $(\bC_{\mathrm{sat}};b_1,\ldots,b_n)$. Now, since $\phi(b_1,\ldots,b_n)$ holds on $\bC_{\mathrm{sat}}$ and $\bA$ is epc for $T$, we find that $\bA \models \phi(a_1,\ldots,a_n)$,
and conclude that $p$ is a maximal ep-type of $T$.

(Backwards.) Take $\bB \models T$, $h:\bA \rightarrow \bB$ a homomorphism, $\bar a$
a tuple of elements of $\bA$,
and $\phi(x_1,\dots,x_n)$ an existential positive formula such that
$\bB \models \phi(h(\bar a))$. Let $p$ be the ep-type of $\bar a$ in $\bA$.
Since $\bB$ is a model of $T$ and $h$ preserves all ep-formulas,
it follows that $T \cup p \cup \{\phi\}$ is satisfiable.
By maximality of $p$, we have that $\phi \in p$, and therefore
$\bA \models \phi(\bar a)$.
\qed
We conclude this section by noting that epc structures are related to the concept of cores, which play such an important role in the classification program for CSPs when the template is finite or $\omega$-categorical.
\begin{prop}
\label{prop:cores}
If $\A$ is epc for $\mathrm{Th}_{\forall^-}(\A)$, then $\A$ is a core. If $\A$ is ep-saturated (or finite) and expanded by all ep-definable relations, then the converse holds also.
\end{prop}
\proof
Suppose $\A$ is epc for $\mathrm{Th}_{\forall^-}(\A)$. Take a homomorphism $h:\A\rightarrow \A$. By epc, for $a_1,\ldots,a_k$ in $A$, if $\A \models R(h(a_1),\ldots,h(a_k))$ or $\A \models h(a_1)=h(a_2)$, then $\A \models R(a_1,\ldots,a_k)$ or $\A \models a_1=a_2$, respectively. It follows that $h$ is an embedding.

Now suppose that $\A$ is ep-saturated (or finite) and expanded by all ep-definable relations. Suppose $\B \models \mathrm{Th}_{\forall^-}(\A)$ and $h:\A\rightarrow \B$ is a homomorphism. Suppose $\B \models \phi(h(\bar a))$, where $\phi(\bar x)$ is an ep-formula and $\bar a$ is a tuple from $A$; we must prove that $\A \models \phi(\bar a)$. 
Let $\langle a_\alpha \rangle_{\alpha < |A|}$ well-order $A$. Consider $(\B;\langle h(a_\alpha) \rangle_{\alpha < |A|})$, by L\"owenheim-Skolem there is an elementarily equivalent $(\B';\langle h(a_\alpha) \rangle_{\alpha < |A|})$ such that $\B'$ is of cardinality no greater than $|A|$, $\B' \models \mathrm{Th}_{\forall^-}(\A)$, $h:\A\rightarrow \B'$ is a homomorphism and $\B' \models \phi(h(\bar a))$.
Since $\B' \models \mathrm{Th}_{\forall^-}(\A)$ there is a homomorphism $g:\B'\rightarrow \A$ by Lemma~\ref{lem:sat-hom} and ep-saturation of $\A$. Therefore $\A \models \phi(g \circ h(\bar a))$ where $g \circ h$ is an endomorphism of $\A$, which must be an embedding since $\A$ is a core. The result $\A \models \phi(\bar a)$ follows as $\A$ is expanded by all ep-definable relations (note that ep-definable relations are not in general preserved by inverse embeddings, but atomic relations are).
\qed
\noindent We note that the additional assumption of being expanded by all ep-definable relations was necessary. Let $\mathbb{Q}^+_0$ be the set $\{q \in \mathbb{Q}:q\geq 0\}$. Then $(\mathbb{Q}^+_0;<)$ is a saturated core but is not epc for $\mathrm{Th}_{\forall^-}(\mathbb{Q}^+_0;<)$.

\subsection{Core theories}
\label{sec:core-theories}
We now introduce the notion of core theories, models of which will be cores. We develop a certain existential-positive model theory, akin to the existential model theory developed in \cite{Simmons72,Simmons76}. Ultimately this will enable us to give a necessary and sufficient condition as to when a template $\bA$ has an equivalent $\bB$ (\mbox{i.e., s.t.} $\mbox{CSP}(\bA)=\mbox{CSP}(\bB)$) such that $\bB$ is a model-complete core. 

We need the concept of a \emph{diagram} of a structure,
in various variants. For a model $\bA$, let $\bA^c$ be the expansion of $\bA$ by $|A|$ constants naming the elements of $\bA$.
Let $\text{diag}_\text{fo}(\bA)$ denote the \emph{elementary diagram of $\bA$}, that is, the set of all fo-sentences true in the structure $\bA^c$. The set of all qf-sentences
that hold on $\bA^c$ is denoted
by $\text{diag}(\bA)$, 
and the set of all universal negative sentences
that hold on $\bA^c$ is denoted
by $\text{diag}_{\forall^-}(\bA)$. 
Finally, 
$\text{diag}_+(\bA)$ denotes the set of all positive qf-sentences  that hold on $\bA^c$.
\begin{lem}[Diagram lemma; Lemma 1.4.2.~in~\cite{Hodges}]\label{lem:diagrams}
Let $\bA$ and $\bB$ be $\tau$-structures, and let $C$ be the set of new constants involved in the expansion $\bA^c$.
Then the following are equivalent.
\begin{enumerate}[\em(1)]
\item There is a $\tau \cup C$-expansion $\bB'$ of $\bB$ \mbox{s.t.} $\bB' \models \text{diag}_+(\bA)$.
\item There is a homomorphism from $\bA$ to $\bB$.
\end{enumerate}
\end{lem}
We will make use of the following well-known preservation theorems.

%
\begin{thm}[Homomorphism Preservation Theorem; see e.g.~Exercise~2 in Section 5.5 of~\cite{Hodges}]
\label{thm:hpt}
Let $T$ be a fo-theory.
A fo-formula $\phi$ is equivalent to an ep-formula modulo $T$ if and only if $\phi$ is preserved
by all homomorphisms between models of $T$.
\end{thm}
\noindent Note that ep and pp are not interchangeable in the previous theorem.
\begin{thm}[Chang-{\L}o\'s-Suszko Theorem; Theorem~6.5.9 in~\cite{HodgesLong} and remarks after the proof]
\label{thm:chains}
Let $T$ be an fo-$\tau$-theory. An fo-$\tau$-formula $\phi$ is equivalent modulo $T$
to a $\forall \exists$-formula if and only if $\phi$ is preserved in unions of chains $(A_i)_{i \leq \kappa}$ whenever
$\bigcup A_i$ and all the $A_i$ are models of $T$.
\end{thm}


\subsubsection{Positive version of Chang-{\L}o\'s-Suszko}
The following fact is called the \emph{existential amalgamation theorem} in~\cite{Hodges}.

\begin{prop}[Theorem 5.4.1 in~\cite{Hodges}]\label{prop:existential-amalgamation}
Let $\bA$ and $\bB$ be $\tau$-structures, $\bar a$ a sequence
of elements of $\bA$, and $h: \bar a \rightarrow \bB$ an embedding
such that every existential sentence true in $(\bB,h(\bar a))$ 
is also true in $(\bA,\bar a)$. Then there exists an elementary 
extension $\bC$ of $\bA$ and an embedding $g: \bB \rightarrow \bC$
such that $g(h(\bar a))=\bar a$. 
\end{prop}

\noindent The existential amalgamation theorem has the following positive variant. We give the proof (it is omitted in~\cite{Hodges}).

\begin{prop}[Theorem 5.4.7 in~\cite{Hodges}]
\label{prop:existential-positive-amalgamation}
Let $\bA$ and $\bB$ be $\tau$-structures, $\bar a$ a sequence
of elements of $\bA$, and $h: \bar a \rightarrow \bB$ an embedding
such that every existential positive sentence true in $(\bB,h(\bar a))$ 
is also true in $(\bA,\bar a)$. Then there exists an elementary 
extension $\bC$ of $\bA$ and a homomorphism 
$g: \bB \rightarrow \bC$
such that $g(h(\bar a))=\bar a$. 
\end{prop}
\proof
Similar to the proof of 5.4.1 in~\cite{Hodges}.
Since $h$ is an embedding, we can replace $\bB$
by an isomorphic copy and assume that $h$ is the identity on $\bar a$.
 Let $\bA'$ be the expansion of $\bA$ by those constant symbols that denote the elements of $\bar a$ in $\bB^c$.
 By Lemma~\ref{lem:diagrams},
it suffices to show that the theory $T := \text{diag}_{\text{fo}}(\bA') \cup \text{diag}_{+}(\bB)$ is consistent.
If $T$ has no model, then by the compactness theorem there is a conjunction $\phi(\bar a, \bar c)$ of finitely many sentences in $\text{diag}_{+}(\bB)$ such that $\bA \models \neg \exists \bar y. \phi(\bar a,\bar y)$. Since $\phi$ is quantifier-free and positive,
the assumptions imply that
$\bB \models \neg \exists \bar y. \phi(\bar a,\bar y)$.
This contradicts that
$\phi(\bar a,\bar c)$ is true in $\bB$.
\qed

\begin{lem}\label{lem:pos-theory-change}
Let $T$ be a theory and $\bA$ a model of the
$\forall\exists^+$-consequences of $T$.
Then $\bA$ can be extended to a model $\bB$ of $T$ such that every 
ep-formula that holds on a tuple $\bar a$ 
in $A \subseteq \bB$ also holds on $\bar a$ in $\bA$.
\end{lem}
\proof
It suffices to prove that
$T \cup \text{diag}(\bA) \cup \text{diag}_{\forall^-}(\bA)$ 
has a model $\bB$. Suppose for contradiction that it were
inconsistent; then by compactness, there exists a finite
subset $U$ of $\text{diag}_{\forall^-}(\bA) \cup  \text{diag}(\bA)$ 
such that $T \cup U$ is inconsistent. 
Let $\phi$ be the conjunction over $U$ where all constants
are existentially quantified. Then $T \cup \{\phi\}$
is inconsistent as well, and $\neg \phi$ is equivalent to a $\forall\exists^+$ sentence, and a consequence
of $T$.
Hence, $\bA \models \neg \phi$, a contradiction.
\qed

The following is a positive version of the Chang-{\L}o\'s-Suszko theorem (Theorem~\ref{thm:chains}). 
We say that an fo-formula $\phi(x_1,\dots,x_n)$ is \emph{preserved in direct limits of models of $T$} if for all sequences  $(\bA_i)_{i<\kappa}$ where all the $\bA_i$ and $\bA := \lim_{i<\kappa} \bA_i$ are models of $T$, and every $n$-tuple $\bar a$ of elements of 
$A$ we have $\bA \models \phi(\bar a)$ whenever
there is for every $i < \kappa$ an $n$-tuple $\bar a^i$ where the $j$-th entry is a representative of the $j$-th entry of $\bar a$, and
$\bA_i \models \phi(\bar a^i)$.

\begin{prop}
\label{prop:pos-chains}
Let $T$ be a $\tau$-theory, and $\Phi$ a set of $\tau$-formulas (in a finite set of free variables). Then the following are equivalent.
\begin{enumerate}[\em(1)]
\item $\Phi$ is modulo $T$ equivalent
to a set of $\forall\exists^+$-formulas;
\item $\Phi$ is preserved in direct limits of sequences of models of $T$;
\item $\Phi$ is preserved in direct limits of countable sequences
of models of $T$.
\end{enumerate}
\end{prop}
\proof
(1) implies (2) follows from Proposition~\ref{prop:direct-product-preservation}.

The implication from (2) to (3) is trivial.
For the 
implication from (3) to (1), assume that $\Phi$ is preserved by direct limits
of sequences $(\bA_i)$ as in the statement of the proposition.
We can assume that $\Phi$ is a set of sentences (by adding constants; see the proof of Theorem~5.4.4. in~\cite{Hodges}).
Let $\Psi$ be the set of all $\forall\exists^+$-sentences
that are consequences of $T \cup \Phi$. We show that 
$T \cup \Psi$ implies $\Phi$, whereupon the result follows. It suffices to show that 
every model of $T \cup \Psi$ is elementary equivalent to a direct limit
of a sequence $(\bB_i)_{i < \omega}$ of models of $T \cup \Phi$ where there are homomorphisms $f_{ij}: {\mathfrak B}_i \rightarrow {\mathfrak B}_j$ with $f_{jk} \circ f_{ij} = f_{ik}$ 
for all $i \leq j \leq k$. 

To construct this sequence, we define an elementary chain of
models $(\bA_i)_{i < \omega}$ of $T \cup \Psi$ such that there are 
\begin{iteMize}{$\bullet$}
\item homomorphisms $f_i: \bA_i \rightarrow \bB_i$, 
with $\bB_i \models T \cup \Phi$, such that for 
every tuple $\bar a_i$ of elements from $\bA_i$ and
every existential positive
formula $\theta$, if $\bB_i \models \theta(f_i(\bar a_i))$,
then $\bA_i \models \theta(\bar a_i)$, and 
\item homomorphisms $g_i: \bB_i \rightarrow \bA_{i+1}$, such that $g_i \circ f_i$ is the identity on $\A_i$.
\end{iteMize}



Let $\bA_0$ be a countable model of $T \cup \Psi$. To  construct the rest of the sequence, suppose that $\bA_i$ has been chosen. 


Since $\bA_0$ is an elementary substructure of $\bA_i$, in particular
all the $\forall\exists^+$-consequences of $T \cup \Phi$ hold in $\bA_i$. By Lemma~\ref{lem:pos-theory-change},
the structure $\bA_i$ 
can be extended to a model $\bB_i$ of $T \cup \Phi$ such that
every ep-sentence that holds
 in $(\bB_i,\bar a_i)$ also holds in $(\bA_i,\bar a_i)$ (\mbox{i.e.} $f_i$ is the identity).
Then by Proposition~\ref{prop:existential-positive-amalgamation},
there are an elementary extension $\bA_{i+1}$ of $\bA_i$ and a homomorphism $g_i: \bB_i \rightarrow \bA_{i+1}$ such that $g_i \circ f_i$ is the identity on $\bA_i$. 

Then $\bC := \bigcup_{i < \omega} \bA_i$ equals $\lim_{i < \omega} \bB_i$,
and by the Tarski-Vaught elementary chain theorem (Theorem~2.5.2 in~\cite{Hodges}) $\bA_0$ is an elementary substructure of $\bC$.
So $\bC$ is a model of $T$, and the direct limit of 
models $\bB_i$ of $T \cup \Phi$, and hence $\bC \models \Phi$.
\qed
\noindent Note that the variant of Proposition~\ref{prop:pos-chains}, in which $\Phi$ is a single formula $\phi$, holds as its corollary, via an application of compactness at the end.

\subsubsection{Model-complete core theories}

A theory $T$ is \emph{model-complete} if any embedding across models of $T$ is elementary, and $T$ is called a \emph{core theory} if every homomorphism between models of $T$ is an embedding. 
The theory of a core need not be a core theory. The integers with successor $(\mathbb{Z};\mathrm{succ})$ is a core, but the homomorphisms from the non-atomic countable models of its theory into $(\mathbb{Z};\mathrm{succ})$ are not even injective (and so are not embeddings).
The following closely follows the proof of Theorem~8.3.1 in~\cite{HodgesLong}.

\begin{thm}\label{thm:mccoretheory}
Let $T$ be an fo-theory over signature $\tau$. 
Then the following are equivalent.
\begin{enumerate}[\em(1)]
\item $T$ is a model-complete core theory.
\item Every model of $T$ is a model epc for $T$.
\item If $\bA,\bB$ are models of $T$ and $h$
is a homomorphism from $\bA$ to $\bB$ then there are an elementary extension $\mathfrak C$ of $\bA$ and an embedding $g$ of $\mathfrak B$ into $\mathfrak C$ such that $g \circ h$ is the identity on $\mathfrak A$. 
\item Every fo-formula is equivalent to an ep-formula modulo $T$.
\end{enumerate}
\end{thm}
\proof
(1) implies (2) is immediate from the definition of epc~models:
if $\bA$ and $\bB$ are models of $T$ and $h: A \rightarrow B$ is 
a homomorphism from $\bA$ to $\bB$, then $h$ must
be an embedding since $T$ is a core theory, and in fact must be
elementary since $T$ is model-complete. Hence, for every tuple
$\bar a$ from $A$ and any ep-formula $\phi$
such that $h(\bar a)$ satisfies $\phi$ we have that $\bar a$ also 
satisfies $\phi$.

(2) implies (3). Assume (2).
Let $\bA$ and $\bB$ be models of $T$, and let
$h$ be a homomorphism from $\bA$ to $\bB$. 
Choose $\bar a$ to
be a vector that enumerates the elements of $\bA$. Since 
$\bA$ is epc for $T$, $h$ is an embedding.
Hence, every existential sentence that holds in $(\bB,h(\bar a'))$ also holds in $(\bA,\bar a')$, for any finite subtuple $\bar a'$ of $\bar a $. 
Proposition~\ref{prop:existential-amalgamation} 
now directly implies (3).

(3) implies (4).
We first claim that if (3) holds, then every homomorphism
between models of $T$ preserves all universal $\tau$-formulas.
For if $h$ is a homomorphism of $\bA$ into $\bB$,
$\bar a$ a tuple from $A$ and $\phi(\bar x)$
a universal $\tau$-formula such that
$\bA \models \phi(\bar a)$, then taking $\bC$ and $g$ as in (3)
we have $\bC \models \phi(g(h(\bar a)))$ and so $\bB \models \phi(h(\bar a))$ since $\phi$ is a universal formula. This proves the claim.
It follows by Theorem~\ref{thm:hpt} that all universal
$\tau$-formulas are equivalent to ep-$\tau$-formulas.

To finally prove (4), let $\phi(\bar x)$ be any fo-$\tau$-formula, \mbox{w.l.o.g.} in prenex normal form.
By a simple induction on the number of quantifier-blocks we can transform $\phi$ to an existential formula, using the fact that the innermost quantifier block
is either existential or universal, and can therefore be transformed
into an existential formula (see Theorem~8.3.1 in~\cite{HodgesLong}).
Finally, existential $\tau$-formulas are preserved by homomorphisms between models of $T$, since such homomorphisms must be embeddings. Hence, the entire formula is even equivalent
to an ep-formula by Theorem~\ref{thm:hpt}.

(4) implies (1). Let $\A$ and $\B$ be models of $T$. Any homomorphism from $\A$ to $\B$ preserves all ep-formulas and therefore all fo-formulas, and hence must be an elementary embedding.
\qed

From this we obtain the following. The proof is similar to the one
for Lindstr\"om's test (see  Theorem 8.3.4 in~\cite{HodgesLong}).

\begin{prop}[Positive version of Lindstr\"om's test]
\label{prop:lindstroem}
Let $T$ be a theory with
signature $\tau$ which
has no finite models but for which there exists a model epc for $T$ of cardinality $\lambda \geq |\tau|$. If $T$ is $\lambda$-categorical, then $T$ is a model-complete core theory.
\end{prop}
\proof[Sketch]
We prove that every model of $T$ is a model epc for $T$ and use
Theorem~\ref{thm:mccoretheory}.
So let $\mathfrak A$ and $\mathfrak B$ be two models of $T$ and
let $h$ be a homomorphism from $\mathfrak A$ to $\mathfrak B$.
Let $\bar a$ be a tuple such that $\bB \models \phi(h(\bar a))$ and
suppose for contradiction that $\bA \not\models \phi(\bar a)$.
Then we can put those two structures into a new 2-sorted structure (comprising $\bA$, $\bB$ and the homomorphism $h$  between them) with fo-theory $T^+$, and 
apply the L\"owenheim-Skolem theorem to produce a countable model
of $T^+$ (where both sorts have the same cardinality since $T$ has no finite models).
By applying L\"owenheim-Skolem again, this time to $T^+$ augmented by sentences expressing a bijection between the two sorts (over a signature expanded by a new function symbol), 
we obtain a two-sorted model of $T^+$
where each sort has cardinality $\lambda$, inducing structures $\mathfrak C$ and $\mathfrak D$, respectively. 
By assumption there exists a model epc for $T$ of
cardinality $\lambda$, with $\lambda \geq |\tau|$,
and by $\lambda$-categoricity $\mathfrak C$ is a model epc for $T$.
This contradicts the fact that we can express in $T^+$ that
$\mathfrak A$ is not a model epc for $T$.
\qed 

\begin{prop}\label{prop:mccoretheory}
Let $T$ be a model-complete core theory. Then $T$ is equivalent
to a $\forall\exists^+$-theory.
\end{prop}
\proof
This is an immediate consequence 
of Proposition~\ref{prop:pos-chains}, since for any sequence
$(\bB_i)_{i < \kappa}$ of models of $T$ with homomorphisms $g_{ij}: \bB_i \rightarrow \bB_j$, the $g_{ij}$ are elementary. By
the Tarski-Vaught theorem on unions of elementary chains (Theorem~2.5.2 in~\cite{Hodges}), we have that $\lim_{i<\kappa} \bB_i \models T$.
\qed


\begin{prop}\label{prop:pos-restr}
A $\lambda$-categorical $\tau$-structure $\bA$, with $\lambda \geq |\tau|$, and epc for the $\forall\exists^+$ restriction of Th$(\bA)$, is a model-complete core iff it has a theory equivalent to a $\forall\exists^+$-theory.
\end{prop}
\proof
The forwards direction follows from Proposition~\ref{prop:mccoretheory}. For the backward direction, we use Proposition~\ref{prop:lindstroem} to derive that $\mathrm{Th}(\bA)$ is a model-complete core theory, whereupon $\bA$ is a model-complete core.
\qed
\begin{prop}
Let $T$ be an $\omega$-categorical theory without finite models (over a finite or countable signature).
Then $T$ is a model-complete core theory if and only if it
is equivalent to a $\forall\exists^+$-theory.
\end{prop}
\proof
The forward direction follows from Proposition~\ref{prop:mccoretheory}. For the backward direction, note that there exists an at most countable epc model for $T$ (which is $\omega$-categorical). We now apply Proposition~\ref{prop:lindstroem} to deduce that $T$ is a model-complete core theory.
\qed

\subsubsection{Core companions}
\begin{defi}
\label{def:cc}
Let $T$ be a fo-$\tau$-theory. Then a $\tau$-theory $U$
is called a \emph{core companion} of $T$ if
\begin{iteMize}{$\bullet$}
\item $U$ is a model-complete core theory;
\item every model of $U$ homomorphically maps to a model of $T$;
\item every model of $T$ homomorphically maps to a model of $U$.
\end{iteMize}
\end{defi}

\begin{prop}
\label{prop:companion-un}
Let $T$ and $T'$ be $\tau$-theories. The following are equivalent.
\begin{enumerate}[\em(1)]
\item Every model of $T$ has a homomorphism to a model of $T'$, and every model of $T'$ has a homomorphism to a model of $T$.
\item $T$ and $T'$ entail the same universal negative sentences.
\end{enumerate}
\end{prop}
\proof
To prove the implication from $(1)$ to $(2)$,
assume $(1)$, and let $\phi$ be a universal negative sentence entailed by $T'$.
Suppose for contradiction that $T$ has a model $\bC$ such that $\bC \models \neg \phi$.
By $(1)$, there is a homomorphism from $\bC$ to a model $\bB$ of $T'$.
Since $\neg \phi$ is equivalent to an existential positive sentence, is preserved by homomorphisms,
and hence we have a contradiction to the assumption that $T'$ entails $\phi$.

For the implication from $(2)$ to $(1)$, assume $(2)$, and let $\bB$ be a model of $T$.
Let $S$ be the ep-theory of $\bB$. We claim that $S \cup T'$ is satisfiable.
If not, then by compactness there is some finite subset
$\{\phi_1,
\dots,\phi_k\}$ of $S$ such that $T'$ entails $(\neg \phi_1\vee \cdots \vee \neg \phi_k)$.
The formula $\neg \phi_1 \vee \cdots \vee \neg \phi_k$ is equivalent to a universal negative sentence $\psi$,
and $T'$ entails $\psi$, so by $(2)$ we have that $T$ entails $\psi$, and hence $\bB \models \psi$.
We have reached a contradiction, since $\bB \models \phi_i$ for all $i \leq k$.
So there indeed exists a model $\bA$ of $S \cup T'$.
The positive variant of the existential amalgamation theorem (Proposition~\ref{prop:existential-positive-amalgamation})
applied to $\bA$ and $\bB$ for the empty sequence $\bar a$ gives a model $\bC$ of $T' \cup S$ and a homomorphism
from $\bB$ to $\bC$.
\qed
The following two propositions are similar to Theorem~8.3.6 in \cite{HodgesLong}.
\begin{prop} \label{prop:uniqueness-core-companion}
Let $T$ be a $\forall\exists^+$-theory 
with signature $\tau$.
If $T$ has a core companion $U$, then $U$ is up to equivalence of theories unique, and is the theory of the class of all models epc for $T$. 
\end{prop}
\proof
We show that the models epc for $T$ are precisely the models of $U$.
First assume that $\mathfrak A$ is a model of $U$. 
Since $U$ is a core companion of $T$, 
there is a homomorphism $e$ from $\bA$ to a model $\bB$ of $T$.
The assumption that $U$ is a core companion of $T$
also implies that there 
exists a homomorphism $f$ from $\bB$ into a model $\bC$ of  $U$.
Then $f \circ e$ is a homomorphism between two models of $U$,
and since $U$ is a model-complete core theory it must be an
elementary embedding. 
This shows in particular that $e$ is an embedding. 
We claim that $\bA$ is a model of the $\forall\exists^+$-theory $T$.
Let $\phi = \forall \bar y. \psi$ be a sentence from $T$ where
$\psi$ is a disjunction of ep and negated atomic $\tau$-formulas, and let $\bar a$ be a tuple from $\bA$.
Since $\bB$ is a model of $T$ and therefore satisfies $\forall \bar y. \psi$, in particular the tuple $e(\bar a)$ satisfies $\psi$. If $e(\bar a)$ satisfies a negated atom in the disjunction $\psi$ then $\bar a$ satisfies $\psi$ as $e$ is an embedding. Otherwise, $e(\bar a)$ satisfies an ep-formula in the disjunction $\psi$ and $f \circ e(\bar a)$ satisfies $\psi$ follows, as $f$ is a homomorphism. Now $\bar a$ satisfies $\psi$ as $f \circ e$ is an elementary embedding.
Since this
holds for all $\bar a$, we have proven that $\bA$ satisfies $\phi$. 

In fact, $\bA$ is a model epc for $T$.
To verify this, let $g$ be a homomorphism from $\bA$ into another model $\bD$ of $T$, $\bar a$ a tuple
from $A$, and $\phi$ an ep-formula with $\bD \models \phi(g(\bar a))$. We have to show that $\bA \models \phi(\bar a)$.
Again, since $U$ is a core companion of $T$
there exists a homomorphism $h$ from $\bD$ into a model $\bE$ of
$U$. Since $U$ is a model-complete core theory, the mapping
$h \circ g$ is elementary. Since $h$ is a homomorphism, 
$\bE \models \phi(h(g(\bar a)))$. Since $h \circ g$ is elementary,
$\bA \models \phi(\bar a)$.

Conversely, we show 
that every model $\bA$ epc for $T$ satisfies $U$.
Since $U$ is a core companion, there is a homomorphism $h$
from $\bA$ to a model $\bB$ of $U$. 
By Proposition~\ref{prop:mccoretheory}, $U$ is equivalent
to a $\forall\exists^+$-theory, 
and thus it suffices to show that $\bA$ satisfies all 
$\forall\exists^+$-consequences $\forall \bar y. \psi(\bar y)$
of  $U$, where $\psi$ is a disjunction of ep
and negated atomic $\tau$-formulas.
Let $\bar a$ be a tuple of elements of $\bA$. We have to show that
$\bA \models \psi(\bar a)$.
Since $\bB \models \forall \bar y. \psi(\bar y)$, 
at least one disjunct $\theta(h(\bar a))$ of $\psi(h(\bar a))$
is true in $\bB$. 
If $\theta$ is a negated atomic formula, then 
$\theta(\bar a)$ is also true in $\bA$ since $h$ is a homomorphism.
If $\theta$ is an ep-formula,
then we deduce $\bC$, a model of $T$, and a homomorphism $g:\bB\rightarrow \bC$ such that $\bC \models \theta(g(h(\bar a)))$. Now $\bA \models \theta(\bar a)$ since $\bA$ is epc for $T$. 
In both cases we can conclude that $\bA \models \psi(\bar a)$.
\qed

\begin{prop} \label{prop:existence-core-companion}
Let $T$ be a $\forall\exists^+$-theory with signature $\tau$.
Then $T$ has a core companion if and only if the class of models epc for $T$ is axiomatizable by a $\tau$-theory.
\end{prop}
\proof
If $T$ has a core companion $U$, 
then Proposition~\ref{prop:uniqueness-core-companion} above implies that $U$ 
axiomatizes the class of models epc for $T$. 

For the converse, 
suppose that the class of models epc for $T$ is the class of all 
models of a $\tau$-theory $U$. Then every model of $U$
is in particular a model of $T$, and every model of $T$
homomorphically maps to a model epc for $T$ (and model of $U$) by Proposition~\ref{prop:existence-epc}. So we only have to verify that $U$ is a model-complete core theory to show that $U$ is the core companion of $T$.

Every
model $\bA$ of $U$ is a model epc for $T$. Since being a model of $U$ implies being a model of $T$, we deduce that $\bA$ is in fact a model epc for $U$.
It follows by the equivalence of (1) and (2) in Theorem~\ref{thm:mccoretheory} that $U$ is a model-complete core theory.
\qed

\begin{cor}
Let $\A$ be a $\tau$-structure, then there exists a $\tau$-structure $\B$ such that $\mathrm{CSP}(\A)=\mathrm{CSP}(\B)$ and $\B$ has a model-complete core theory iff the class of models epc for the $\forall\exists^+$-theory of $\A$ is axiomatizable by a $\tau$-theory.
\end{cor}
\proof (Forwards.) If such a $\B$ exists then Th$(\B)$ must be a model-complete core theory and Th$(\B)$ is the core companion of Th$(\A)$. The result follows from Proposition~\ref{prop:existence-core-companion}.

(Backwards.) If the class of models epc for the $\forall\exists^+$-theory of $\A$ is axiomatizable, then the $\forall\exists^+$-theory of $\A$ has a core companion $U$. Let $\B$ be a model of $U$. It follows from Definition~\ref{def:cc} that $\B$ is a model-complete core and that $\A$ and $\B$ are homomorphically equivalent, hence $\mathrm{CSP}(\A)=\mathrm{CSP}(\B)$.
\qed
\noindent Note that $\A$ and $\A'$ may generate the same CSP, i.e. share the same ep-theory, whilst having distinct $\forall\exists^+$-theories. The previous proof shows us that ep-equivalence does not affect the question as to whether or not the class of models epc for the $\forall\exists^+$-theory of $\A$ is axiomatizable.

\subsection{Equivalent $\omega$-categorical templates}
\label{sec:omega-cat}

We are now in a position to establish precisely when a template $\bA$ has an equivalent $\bB$ (\mbox{i.e., s.t.} $\mbox{CSP}(\bA)=\mbox{CSP}(\bB)$) such that $\bB$ is $\omega$-categorical. At the same time we reprove the result of \cite{Cores-journal} that every $\omega$-categorical template is equivalent to a unique template that is a model-complete core. The result in \cite{Cores-journal} was obtained using ad-hoc arguments that poorly reflected the aspects of our programme for the existential-positive that are similar to the programme for the existential in \cite{Simmons72,Simmons76}.

A structure is \emph{homogeneous} (sometimes called \emph{ultrahomogeneous} \cite{HodgesLong}) if every finite partial automorphism can be extended to a full automorphism. 
\begin{lem}\label{lem:cat-via-homogen}
Let $\A$ be a countable homogeneous structure such that for each $k$ only a finite number of distinct $k$-ary relations
can be defined by atomic formulas. Then $\A$ is $\omega$-categorical.
\end{lem}
\proof
By homogeneity of $\A$, the atomic formulas that hold on the elements of $t$ in $\A$ determine the orbit of $t$ under Aut$(\A)$. Since there are
only finitely many such atomic formulas, it follows that there are finitely many orbits of $k$-tuples in Aut$(\A)$. The claim follows by Theorem~\ref{thm:ryll}.
\qed
A theory $T$ is said to have the \emph{Joint Homomophism Property} (JHP) if, for all models $\bA$ and $\bB$ of $T$, there exists a model $\bC$ of $T$ such that both $\bA$ and $\bB$ map homomorphically into $\bC$.
\begin{prop}\label{prop:jhp}
For any theory $T$, the following are equivalent.
\begin{enumerate}[\em(1)]
\item For all ep-$\tau$-sentences $\phi_1$ and
$\phi_2$, if $T \cup \{\phi_1\}$ is satisfiable and $T \cup \{\phi_2\}$ is satisfiable
then $T \cup \{\phi_1, \phi_2\}$ is satisfiable as well.
\item $T$ has the JHP.
\end{enumerate}
\end{prop}
\proof
Assume $(1)$, and let $\tau$ be the signature of $T$.
Let $\bA$ and $\bB$ be models of $T$.
We claim that the theory $T' := T \cup \text{diag}_+(\bA) \cup \text{diag}_+(\bB)$ is satisfiable.
By compactness, it suffices to show that every finite subset $S$ of $T'$
is satisfiable. Let $S_1 := S \cap \text{diag}_+(\bA)$ and $S_2 := S \cap \text{diag}_+(\bB)$. By forming a finite conjunction, we see that $S_1$ and $S_2$ are logically
equivalent to single sentences $\phi_1$ and $\phi_2$, respectively. Let $\phi'_1$ and $\phi'_2$ be the $\tau$-sentences corresponding to $\phi_1$ and $\phi_2$, respectively, in which the new constant symbols are existentially quantified out.
Certainly $T \cup \{\phi'_1\}$ and $T \cup \{\phi'_2\}$ are satisfiable since $\bA$ and $\bB$ are models of $T$ and therefore satisfy all sentences from $T$.
By $(1)$,  the theory $T \cup \{\phi'_1, \phi'_2\}$ is satisfiable as well.
Therefore the claim is true, and there exists a model $\bC'$ of $T'$.
Let $\bC$ be the $\tau$-reduct of $\bC'$.
Finally, Lemma~\ref{lem:diagrams} asserts the existence of a homomorphism from $\bA$ to $\bC$, and from $\bB$ to $\bC$, which proves $(2)$.

For the implication from $(2)$ to $(1)$, suppose that $T$ is a $\tau$-theory with the JHP, and that
$\phi_1$ and $\phi_2$ are ep-sentences such that $T \cup \{\phi_1\}$ has a model $\bA$
and $T \cup \{\phi_2\}$ has a model $\bB$. By $(2)$, there exists a model of $T$ such that $\bA$ and $\bB$ homomorphically
map to $\bC$. Then $\bC$ clearly satisfies $T \cup \{\phi_1,\phi_2\}$.
\qed

For a satisfiable theory $T$, let $\sim_n^T$ be the equivalence relation defined on ep-formulas with $n$ free variables
$x_1,\dots,x_n$ as follows.
For two such formulas $\phi_1$ and $\phi_2$, 
let $\phi_1 \sim^T_n \phi_2$
iff for all ep-formulas $\psi$ with free variables $x_1,\dots,x_n$
we have that $\{\phi_1, \psi\} \cup T$ is satisfiable if
and only if $\{\phi_2,\psi \} \cup T$ is satisfiable. By proving that a model epc for a certain type of theory is in fact $\omega$-categorical, we will derive the following.
\begin{thm}
\label{thm:omega-cat}
Let $T$ be a theory with the JHP.
Then the following are equivalent.
\begin{enumerate}[\em(i)]
\item $T$ has a core companion, unique up to equivalence of theories, that is either $\omega$-categori\-cal or the theory of a finite structure.
\item $\sim^T_n$ has finite index for each $n$.
\item $T$ has finitely many maximal ep-$n$-types for each $n$.
\item There is a finite or $\omega$-categorical model $\A$ that is a model-complete core, and which satisfies an existential positive sentence $\phi$ iff $T \cup \{\phi\}$ is satisfiable.
\end{enumerate}
\end{thm}
\proof
$(i) \Rightarrow (ii)$.
Let $U$ be the core companion of $T$, since they entail the same universal negative sentences, we can deduce -- for ep-formulas $\psi$ -- that $U \cup \{\psi\}$ is satisfiable if and only if $T \cup \{\psi\}$ is satisfiable. It follows that the indices of $\sim_n^U$ and $\sim_n^T$ coincide. 

Let $A$ be a finite, or the countable, model of $U$. For a proof by contraposition, assume $\sim^U_n$ has infinite index for some $n$. Let $\phi_1$ and $\phi_2$ be two 
ep-formulas from different equivalence classes of $\sim^U_n$. Hence, there is an ep-formula
$\phi_3$ with free variables $x_1,\dots,x_n$ 
such that exactly one of the two formulas $\phi_1 \wedge \phi_3$ and $\phi_2 \wedge \phi_3$
is satisfiable relative to $U$.
This shows that $\phi_1$ and $\phi_2$ define over $\A$ distinct
relations, and therefore that $\A$ can not be $\omega$-categorical (as Theorem~\ref{thm:ryll} asserts that it has only a finite number of inequivalent first-order definable relations of arity $n$, and in particular only a finite number of inequivalent ep-definable relations of arity $n$).

$(ii) \Rightarrow (iii)$.
We show that every maximal ep-$n$-type $p$ is determined completely by the $\sim^T_n$ equivalence classes of the ep-formulas contained in $p$. Since there are finitely many such classes, the result follows. Let $p$ and $q$ be maximal ep-$n$-types \mbox{s.t.} for every $\phi_1 \in p$, exists $\phi'_1 \in q$ s.t. $\phi_1\sim^T_n \phi'_1$ and for every $\phi_2 \in q$, exists $\phi'_2 \in p$ \mbox{s.t.} $\phi_2\sim^T_n \phi'_2$. We aim to prove that $p=q$. If not then there exists, \mbox{w.l.o.g.}, $\psi \in p$ \mbox{s.t.} $\psi \notin q$. Clearly, $T \cup p \cup \{\psi\}$ is satisfiable, and, since $q$ is maximal, $T \cup q \cup \{\psi\}$ is not satisfiable. By compactness $T \cup \{\theta_q, \psi\}$ is not satisfiable for some finite conjunction $\theta_q$ of formulas from $q$. Now, $\theta_q \in q$ by maximality and there exists by assumption $\theta'_q \in p$ \mbox{s.t.} $\theta_q \sim^T_n \theta'_q$. By definition of $\sim^T_n$ we deduce $T \cup \{ \theta'_q, \psi\}$ satisfiable iff $T \cup \{ \theta_q, \psi\}$ satisfiable. Since the latter is not satisfiable, we deduce that neither is the former, which yields the contradiction that $T \cup p \cup \{\psi\}$ is not satisfiable.

$(iii) \Rightarrow (iv)$. We aim to show that 
\[ S:=\{ \phi : \phi \mbox{ ep-formula consistent with $T$} \} \cup \{ \neg \phi : \phi \mbox{ ep-formula not consistent with $T$} \}\]
has a finite or $\omega$-categorical model that is a model-complete core.
Since $T$ has the JHP, it follows by induction that $S$ is finitely satisfiable, and thence by compactness that $S$ is satisfiable. We claim that $S$ has finitely many maximal ep-$n$-types for each $n$. It suffices to prove that if $p$ is a maximal ep-$n$-type of $S$ then it is also a maximal ep-$n$-type of $T$. If $\phi$ is an ep-formula, then $T \cup p \cup \{\phi\}$ being satisfiable implies $S \cup p \cup \{\phi\}$ is satisfiable (by induction through JHP and compactness), and the claim follows.  
Let the number of maximal ep-$n$-types of $S$, $\mu_n$, be finite for all $n$. We will show that $S$ has an $\omega$-categorical model. We consider the signature $\tau'$, which is the expansion of $\tau$ by $\mu_n$ relations of each arity $n$, corresponding to the maximal ep-$n$-types of $S$. 
Any model of $S$ has a canonical (unique) expansion to a $\tau'$-structure (by the new relation symbols labeling tuples that attain their type). Consider the canonical $\tau'$-expansion $\A'$ of a countable or finite $\tau$-model $\A$ epc for $S$, guaranteed to exist by Proposition~\ref{prop:existence-epc}. We will shortly prove that $\A'$ is homogeneous. From this it will follow that $\A'$ is $\omega$-categorical by Lemma~\ref{lem:cat-via-homogen} (there is only a finite number of inequivalent atomic formulas of each arity $n$, since the collapsings of formulas with higher arity are themselves already $n$-ary atoms of $\tau'$), whereupon $\omega$-categoricity is inherited by its $\tau$-reduct $\A$. Since $\A$ is epc for $S$, it can be seen to also be epc for $\mathrm{Th}_{\forall^-}(\A)$, and it follows from Proposition~\ref{prop:cores} that $\A$ is a core. Homogeneity enforces that each fo-formula is equivalent modulo $\mathrm{Th}(\A')$ to a Boolean combination of ep-formulas. However, because we expand by a finite number of maximal ep-$n$-types, each complement of en ep-$n$-type is again an ep-$n$-type, and so the two coincide. It follows that each fo-formula is equivalent modulo $\mathrm{Th}(\A')$ to an ep-formula, and model-completeness of $\mathrm{Th}(\A')$ follows from Theorem~\ref{thm:mccoretheory}. It follows that $\A \models \mathrm{Th}(\A')$ is a model-complete core.

It remains to prove that $\A'$ is homogeneous. An ep-formula $\phi(\overline{x})$ is said to isolate a maximal ep-$n$-type $p(\overline{x})$ of $S$, if $p$ is the only maximal ep-$n$-type of $S$ of which $\phi$ is a member. If there is only a finite number of maximal ep-$n$-types of $S$, then it follows that each has an isolating formula (\mbox{i.e.} a formula that is in that maximal ep-$n$-type but in no other maximal ep-$n$-type). Let $f:(a_1,\ldots,a_m)\mapsto (b_1,\ldots,b_m)$ be a partial automorphism of $\A'$ (in the signature $\tau'$). Let $a'$ be an arbitrary element of $A'$. Consider the ep-$n$-types $p(x_1,\ldots,x_m)$ of $(a_1,\ldots,a_m)$ and $q(x_1,\ldots,x_m,y)$ of $(a_1,\ldots,a_m,a')$ in $\A$. By Proposition~\ref{prop:epc-types}, each of these types is maximal, and is isolated by the ep-formulas $\theta_p(x_1,\ldots,x_m)$ and $\theta_q(x_1,\ldots,x_m,y)$, respectively. Furthermore, the type of $(b_1,\ldots,b_m)$ in $\A$ is $p$ (as the partial automorphism of $\A'$ respects the signature $\tau'$). But now, since $\exists y.\theta_q(x_1,\ldots,x_m,y) \in p$ (by maximality), we may deduce a $b'$ s.t. $\A' \models \theta_q(b_1,\ldots,b_m,b')$ and consequently $\A' \models q(b_1,\ldots,b_m,b')$. It follows that $f':(a_1,\ldots,a_m,a')\mapsto (b_1,\ldots,b_m,b')$ is a partial automorphism of $\A'$ (in the signature $\tau'$). A simple back-and-forth argument shows that we may extend to an automorphism of $\A'$, and the result follows.

$(iv) \Rightarrow (i)$.
Let $\bA$ be the finite or $\omega$-categorical structure that is a model-complete core. We claim Th$(\bA)$ is a core companion of $T$. This follows from the fact that $T$ and Th$(\bA)$ entail the same universal negative sentences, via Proposition~\ref{prop:companion-un}. Observe that Th$(\bA)$ is equivalent to a $\forall\exists^+$ theory, by Proposition~\ref{prop:mccoretheory}. Being a model-complete core theory follows immediately if $\bA$ is finite, and via Proposition~\ref{prop:lindstroem} if $\bA$ is $\omega$-categorical. Uniqueness follows from Proposition~\ref{prop:uniqueness-core-companion}, since the core companion of Th$(\bA)$ is, up to equivalence of theories, unique (and this implies that Th$(\bA)$ is already the unique core companion of $T$, up to equivalence of theories).
\qed

We end this section by drawing comparison between Theorem~\ref{thm:omega-cat} and a variant that is existential (but not existential positive). A theory $T$ is said to have the \emph{Joint Embedding Property} (JEP) if, for all models $\bA$ and $\bB$ of $T$, there exists a model $\bC$ of $T$ such that both $\bA$ and $\bB$ embed into $\bC$. A \emph{companion} to a theory $T$ is a theory $T'$ \mbox{s.t.} every model of $T'$ can be embedded into a model of $T$ and vice-versa. A \emph{model companion} is a companion that is model-complete. A $\exists_1$-type is defined exactly as an ep-type but with existential (as opposed to existential positive) formulas.
\begin{thm}[\cite{Simmons76}]
\label{thm:simmons}
Let $T$ be a theory with the JEP.
Then the following are equivalent.\footnote{Note that Simmons uses a stronger notion of model companion than we have given here, but the statement can be seen to hold nonetheless.}
\begin{iteMize}{$\bullet$}
\item $T$ has a model companion that is either $\omega$-categorical or the theory of a finite structure.
\item $T$ has finitely many maximal $\exists_1$-$n$-types for each $n$.
\end{iteMize}
\end{thm}
\noindent We also draw the reader's attention to the similarity between our Proposition~\ref{prop:epc-types} and Theorem 1.2(b) in \cite{Simmons76}.

\section{Applications}

We give a series of applications of our results in the study of the structure and complexity of CSPs.

\subsection{Essentially unary polymorphisms}

We will begin by demonstrating that the power of infinitary polymorphisms can be greatly limited. The forthcoming three lemmas are well-known for finite domains (also for $\omega$-categorical structures). They require a little care in the infinite case.

A function $f:A^\alpha \rightarrow A$ is \emph{essentially unary} if there exist a $\beta < \alpha$ and $g:A \rightarrow A$ such that, for all $\overline{x} \in A^\alpha$, $f(\overline{x})=g(x_\beta)$.
For $\overline{x}, \overline{w} \in A^\alpha$, and $X \subseteq \alpha$, let $\overline{x}[\overline{x}_X/\overline{w}_X]$ be the tuple $\overline{x}$ with each entry $x_\beta$, where $\beta \in X$, substituted by $w_\beta$.
\begin{lem}
\label{lem:ess-unary-first}
A function $f:A^\alpha \rightarrow A$ is not essentially unary iff there exist two non-empty and disjoint $X,Y \subseteq \alpha$, such that both
\begin{iteMize}{$\bullet$}
\item exist $\overline{x}, \overline{w},\overline{w}'\in A^\alpha$ s.t. $f(\overline{x}[\overline{x}_X/\overline{w}_X])\neq f(\overline{x}[\overline{x}_X/\overline{w}'_X])$, and
\item exist $\overline{y}, \overline{z},\overline{z}'\in A^\alpha$ s.t. $f(\overline{y}[\overline{y}_Y/\overline{z}_Y])\neq f(\overline{y}[\overline{y}_Y/\overline{z}'_Y])$.
\end{iteMize}
\end{lem}
\proof We will benefit from the following local definition. A set $Z \subseteq \alpha$ is termed \emph{good} if the following holds: for all $\overline{x}, \overline{w},\overline{w}'\in A^\alpha$ we have $f(\overline{x}[\overline{x}_Z/\overline{w}_Z])= f(\overline{x}[\overline{x}_Z/\overline{w}'_Z])$. If $Z$ is not good, then we term it \emph{bad}. Note that good sets are closed under union; i.e., if $X$ and $Y$ are both good, then so is $X \cup Y$. The contrapositive of the lemma is the assertion that $f$ is essentially unary iff, for any two non-empty and disjoint $X,Y \subseteq \alpha$, at least one of $X$ and $Y$ is good. 

(Backwards.) By contraposition. If $f$ is essentially unary, then let $\beta$ and $g$ be s.t. $f(\overline{x})=g(x_\beta)$. Now, take any two non-empty and disjoint $X,Y \subseteq \alpha$. At least one does not contain $\beta$, and it must be a good set.

(Forwards.) By contraposition. Assume that, for any two non-empty and disjoint $X,Y \subseteq \alpha$, at least one of $X$ and $Y$ is good. If there are no bad subsets of $\alpha$, i.e. $f$ is constant, then clearly $f$ is essentially unary. Assume the existence of some bad set. We will derive the existence of a bad set of cardinality $1$; for otherwise let $Z$ be a minimal bad set (under the total lexicographical order on the $0-1$ characteristic sequence of length $\alpha$) of cardinality greater than $1$. Let $Z_1$ and $Z_2$ be a non-trivial partition of $Z$. At least one of $Z_1$ and $Z_2$ must be good, by assumption. Hence the other must be bad (as good sets are closed under union, and $Z:=Z_1 \cup Z_2$ is bad), contradicting minimality of $Z$. Let $Z=\{\beta\}$ be a minimal bad set. Set 
\[ g(x_\beta):=f({x_\beta}^\alpha)=f(x_\beta,x_\beta,\ldots),\]
i.e. each variable $x_\gamma$, $\gamma \leq \alpha$, is substituted by $x_\beta$ (of course the choice of $x_\beta$ as the variable here is not important). That $f(\overline{x})=g(x_\beta)$ now follows from $\alpha \setminus \{\beta\}$ being a good set. 
\qed
\begin{lem}
\label{lem:pp-equals-ep}
For all $\A$, $(x=y \vee u=v) \in \langle \A \rangle_{\mathrm{pp}}$ iff $\langle \A \rangle_{\mathrm{pp}} = \langle \A \rangle_{\mathrm{ep}}$.
\end{lem}
\proof
The backward direction is trivial. We prove the forward direction. Our proof will be by simulation of the binary $\vee$. Take $\phi \in \langle \A \rangle_{\mathrm{ep}}$ in prenex form; we will recursively remove disjunctions of the form
\[ \psi_1(x_1,\ldots,x_n,y_1,\ldots,y_p) \vee \psi_2(x_1,\ldots,x_n,z_1,\ldots,z_q).\]
We may assume that each of $\psi_1$ and $\psi_2$ is alone satisfiable, for otherwise their disjunction is logically equivalent to just one of them. We will introduce new variables $x'_1,\ldots,x'_n,y'_1,\ldots,y'_p$ and $x''_1,\ldots,x''_n,z'_1,\ldots,z'_q$. It follows from \cite{ecsps} that there is a $\theta \in \langle (A;x=y \vee u=v) \rangle_{\mathrm{pp}}$ such that $\theta \equiv$
\[
\begin{array}{ll}
 (x'_1=x_1 \wedge \ldots \wedge x'_k=x_k \wedge y'_1=y_1 \wedge \ldots \wedge y'_p=y_p) & \vee \\
 (x''_1=x_1 \wedge \ldots \wedge x''_k=x_k \wedge z'_1=z_1 \wedge \ldots \wedge z'_q=z_q).&
\end{array}
\]
The disjunct $\psi_1 \vee \psi_2$ should be replaced with the following, in which the existential quantifiers should be read as all coming before the conjunction.
\[
\begin{array}{lr}
\exists x'_1,\ldots,x'_n,y'_1,\ldots,y'_p. & \psi_1(x'_1,\ldots,x'_n,y'_1,\ldots,y'_p) \wedge \\
\exists x''_1,\ldots,x''_n,z'_1,\ldots,z'_p. & \psi_2(x''_1,\ldots,x''_n,z'_1,\ldots,z'_q) \wedge \\
& \theta(x_1,\ldots,x_n,y_1,\ldots,y_p, \\
& x'_1,\ldots,x'_n,y'_1,\ldots,y'_p, \\
& x''_1,\ldots,x''_n,z'_1,\ldots,z'_q)
\end{array}
\]
\qed
\begin{lem}
\label{lem:essentially-unary}
Let $\A$ be such that $(u=v \vee x=y) \in \langle \A \rangle_{\mathrm{pp}}$. Then all (finitary and infinitary) polymorphisms of $\A$ are essentially unary.
\end{lem}
\proof
Let $P_4:=(u=v \vee x=y) \in \langle \A \rangle_{\mathrm{pp}}$. It follows from Lemma~\ref{lem:pp+} that $P_4$ must be preserved by the polymorphisms of $\A$. Suppose for contradiction that $\A$ has a polymorphism $f:M^\alpha \rightarrow M$ that is not essentially unary. From Lemma~\ref{lem:ess-unary-first}, we deduce non-empty and disjoint $X,Y \subseteq \alpha$, s.t. there exist $\overline{x}, \overline{w},\overline{w}'\in A^\alpha$ with $f(\overline{x}[\overline{x}_X/\overline{w}_X])\neq f(\overline{x}[\overline{x}_X/\overline{w}'_X])$ and $\overline{y}, \overline{z},\overline{z}'\in A^\alpha$ with $f(\overline{y}[\overline{y}_Y/\overline{z}_Y])\neq f(\overline{y}[\overline{y}_Y/\overline{z}'_Y])$. But, for each $\beta \in \alpha$, 
\[ P_4(\overline{x}[\overline{x}_X/\overline{w}_X]_\beta, \overline{x}[\overline{x}_X/\overline{w}'_X]_\beta, \overline{y}[\overline{y}_Y/\overline{z}_Y]_\beta, \overline{y}[\overline{y}_Y/\overline{z}'_Y]_\beta) \]
holds, by disjointness of $X$ and $Y$, while
\[ P_4(f(\overline{x}[\overline{x}_X/\overline{w}_X]), f(\overline{x}[\overline{x}_X/\overline{w}'_X]), f(\overline{y}[\overline{y}_Y/\overline{z}_Y]), f(\overline{y}[\overline{y}_Y/\overline{z}'_Y]) ) \]
does not.
\qed
\begin{prop}
\label{cor:global-unary}
For all structures $\A$, $\langle \A \rangle_\mathrm{pp} = \langle \A \rangle_\mathrm{ep}$ iff all $\omega$-polymorphisms of all elementary extensions of $\A$ are essentially unary.
\end{prop}
\proof
(Forwards.) If $\langle \A \rangle_\mathrm{pp} = \langle \A \rangle_\mathrm{ep}$ then $(u=v \vee x=y) \in \langle \A \rangle_\mathrm{pp}$, and so $(u=v \vee x=y) \in \langle \A' \rangle_\mathrm{pp}$ for all $\A' \succeq \A$. The result follows from Lemma~\ref{lem:essentially-unary}.

(Backwards.) If all $\omega$-polymorphisms of all elementary extensions of $\A$ are essentially unary, then in particular this is true of the monster elementary extension $\M$ built as in Lemma~\ref{lem:monster-model}. It follows from Theorem~\ref{thm:inv-pol-inf} that $(u=v \vee x=y) \in \langle \M \rangle_\mathrm{pp}$, which gives $\langle \M \rangle_\mathrm{pp} = \langle \M \rangle_\mathrm{ep}$ by Lemma~\ref{lem:pp-equals-ep}. The result $\langle \A \rangle_\mathrm{pp} = \langle \A \rangle_\mathrm{ep}$ follows since $\A \preceq \M$. 
\qed
\noindent We are able to prove that the stipulation of elementary extension in Proposition~\ref{cor:global-unary} is necessary, by exhibiting a structure whose $\omega$-polymorphisms include only projections but for which $(x=y \vee u=v)$ is not pp-definable. In the following, $+$ should be read as a ternary relation.
\begin{lem}
\label{lem:Q+1neq}
The only $\omega$-polymorphisms of $(\mathbb{Q};+,1,\neq)$ are projections.
\end{lem}
\proof
We give the proof for polymorphisms of arity $\omega$, but the argument works just as well for any infinite or finite arity. A function $f:D^\omega \rightarrow D$ is \emph{idempotent} if $f(d,d,\ldots)=d$, for all $d \in D$. It is \emph{conservative} if it further satisfies $f(d_1,d_2,\ldots) \in \{d_1,d_2,\ldots\}$, for all $d_1,d_2,\ldots \in D$. Let $f:\mathbb{Q}^\omega \rightarrow \mathbb{Q}$ be a polymorphism of $(\mathbb{Q};+,1,\neq)$. It is clear that $f$ is idempotent as the only endomorphism of $(\mathbb{Q};+,1)$ is the identity. Further, by preservation of $\neq$, it is easy to see that $f$ must be conservative.
Consider $\{0,1\}^\omega$ with the total lexicographical ordering induced by $0<1$. Choose some minimal $\langle z_\lambda \rangle_{\lambda < \omega} \in \{0,1\}^\omega$ s.t. $f(\langle z_\lambda \rangle_{\lambda < \omega})=1$ (since $f(1,1,\ldots)=1$, such a $\langle z_\lambda \rangle_{\lambda < \omega}$ exists). If $\langle z_\lambda \rangle_{\lambda < \omega}$ had more than one index that is a $1$, then there would exist $\langle z'_\lambda \rangle_{\lambda < \omega}$ and $\langle z''_\lambda \rangle_{\lambda < \omega}$ s.t. $\langle z'_\lambda \rangle_{\lambda < \omega}, \langle z''_\lambda \rangle_{\lambda < \omega} < \langle z_\lambda \rangle_{\lambda < \omega}$ and $\langle z'_\lambda \rangle_{\lambda < \omega} + \langle z''_\lambda \rangle_{\lambda < \omega} = \langle z_\lambda \rangle_{\lambda < \omega}$ and so, by preservation of $+$, one of $\langle z'_\lambda \rangle_{\lambda < \omega}, \langle z''_\lambda \rangle_{\lambda < \omega} = 1$, contradicting minimality of $\langle z_\lambda \rangle_{\lambda < \omega}$. So, for some $i$, $\langle z_\lambda \rangle_{\lambda < \omega}$ is of the form
\[ (0, \ldots, 0, \stackrel{\mbox{$i$th position}}{\overbrace{1}},0,\ldots ). \]
By preservation of $+$, it follows, for each $q \in \mathbb{Q}$, that $f(q\cdot \langle z_\lambda \rangle_{\lambda < \omega})=q$. 

Firstly, we consider $\langle x_\lambda \rangle_{\lambda < \omega} \in \mathbb{Q}^\omega$ s.t. $q \notin \{ x_\lambda : \lambda < \omega\} \neq \mathbb{Q}$. If $f(\langle x_\lambda \rangle_{\lambda < \omega}) = p \neq x_i$, then, by preservation of $+$, 
\[ f(\langle x_\lambda \rangle_{\lambda < \omega} + (q-p) \langle z_\lambda \rangle_{\lambda < \omega})= f(\langle x_\lambda \rangle_{\lambda < \omega}) + (q-p) f(\langle z_\lambda \rangle_{\lambda < \omega}) = q. \]
But this violates conservativity of $f$ as $q$ does not appear in $\langle x_\lambda \rangle_{\lambda < \omega} + (q-p) \langle z_\lambda \rangle_{\lambda < \omega}$ (since $p \neq x_i$).

Finally, we take an arbitrary $\langle x_\lambda \rangle_{\lambda < \omega} \in \mathbb{Q}^\omega$. Consider the set $\Lambda := \{\lambda : x_\lambda = 1\}$ and $\langle x'_\lambda \rangle_{\lambda < \omega}$ and $\langle x''_\lambda \rangle_{\lambda < \omega}$ obtained according to $x'_\lambda=x_\lambda$, if $\lambda \notin \Lambda$, and $=0$ otherwise; and $x''_\lambda=1$, if $\lambda \in \Lambda$, and $=0$ otherwise. Clearly $\langle x_\lambda \rangle_{\lambda < \omega} = \langle x'_\lambda \rangle_{\lambda < \omega} + \langle x''_\lambda \rangle_{\lambda < \omega}$, and $\langle x'_\lambda \rangle_{\lambda < \omega}$ and $\langle x''_\lambda \rangle_{\lambda < \omega}$ satisfy the condition of the previous paragraph, i.e. that neither $\{ x'_\lambda : \lambda < \omega\}$ nor $\{ x''_\lambda : \lambda < \omega\}$ is $\mathbb{Q}$. The result follows by preservation of $+$.   
\qed
\noindent It follows that $(x=y \vee u=v) \in \Inv(\Pol^\omega(\mathbb{Q};+,1,\neq))$, though $(x=y \vee u=v)$ is not pp-definable in $(\mathbb{Q};+,1,\neq)$ since if it were we could also derive $(x=y \vee u=v) \in \langle(\mathbb{R};+,1,\neq)\rangle_{\mathrm{pp}}$ (since $(\mathbb{R};+,1,\neq)$ and $(\mathbb{Q};+,1,\neq)$ share the same theory). This would contradict Lemma~\ref{lem:essentially-unary} as $(\mathbb{R};+,1,\neq)$ has polymorphisms that are not essentially unary: indeed, there is an isomorphism between $(\mathbb{R};+,1)^2$ and $(\mathbb{R};+,1)$ (that we shall use again shortly), which gives a bijective homomorphism from $(\mathbb{R};+,1,\neq)^2$ to $(\mathbb{R};+,1,\neq)$.

The following definition comes from \cite{BodirskyHermannRichoux}. For a structure $\A$ and an ep-sentence $\phi$, we generate the boolean sentence $F_\A(\phi)$ by removing all existential quantifiers and replacing each atom $R(x_1,\ldots,x_k)$, where $R^\A$ is empty, with \textit{false}, and replacing all other atoms with \textit{true}. $\A$ is said to be \emph{locally refutable} if for every ep-sentence $\phi$, $\A \models \phi$ iff $F_\A(\phi)$ is true. 
\begin{prop}
\label{prop:not-locally-refutable}
Let $\A$ be a structure that is not locally refutable and for which all $\omega$-polymorphisms in all elementary extensions are essentially unary. Then CSP($\A$) is NP-hard.
\end{prop}
\proof
It is proved in \cite{BodirskyHermannRichoux} that the evaluation of ep-sentences on $\A$ is NP-hard. The result now follows from Lemma~\ref{lem:pp-equals-ep} (note that the recursive removal of disjunction induces a polynomial time reduction).
\qed

\subsection{First-order definable CSPs}

Recall $\phi[\B]$ to be the canonical query of $\B$. CSP$(\A)$ is said to be \emph{first-order definable} if there is an fo-sentence $\psi_\A$ such that, for all finite $\B$, $\A \models \phi[\B]$ (i.e. $\phi[\B] \in \mbox{CSP}(\A)$) iff $\B \models \psi_\A$. The following definition comes from \cite{LLT}.
The \emph{one-tolerant $n$-th power} $^1\mathfrak A^n$ 
of a $\tau$-structure $\mathfrak A$ is the $\tau$-structure with domain $A^n$
where a $k$-ary $R \in \tau$ denotes the relation consisting of all those $k$-tuples $((a_1^1,\dots,a_1^n),\dots,$
$(a_k^1,\dots,a_k^n))$ such that $$|\{j : \; (a_1^j,\dots,a_k^j) \in R^\mathfrak A\} | \geq n-1\; .$$
For $n \geq 3$, an $n$-ary polymorphism $f$ of $\mathfrak A$ is called a \emph{$1$-tolerant} polymorphism 
if $f$ is a homomorphism from $^1\mathfrak A^n$ to $\mathfrak A$.
The following is our analog of the result from \cite{LLT}.
\begin{thm}\label{thm:fo}
Let $\mathfrak A$ be a monster elementary extension (as constructed as in Lemma~\ref{lem:monster-model}) on a finite signature.
Then CSP$(\mathfrak A)$ is first-order definable if and only if
$\mathfrak A$ has a $1$-tolerant polymorphism.
\end{thm}
\proof
\textbf{Claim 1}. 
Let $\M$ be a monster extension. If all finite substructures $\C$ of $^1\M^{n+1}$ map homomorphically to $\M$, then $^1\M^{n+1}$ maps homomorphically to $\M$. 

\textit{Proof of Claim 1.} We note, for structures $\A$ and $\B$, that if $\A$ and $\B$ are elementarily equivalent, then so are $^1\A^{n+1}$ and $^1\B^{n+1}$ (as $^1\A^{n+1}$ is fo-definable in $\A$). The assumption of the claim may be restated as that all pp-$\tau$-sentences true in $^1\M^{n+1}$ are true in $\M$. This is clearly true by assumption that all finite substructures $\C$ map homomorphically to $\M$.

A finite $\tau$-structure $\mathfrak C$ is an \emph{obstruction} for the 
$\tau$-structure $\mathfrak A$ if there is no homomorphism from $\mathfrak C$ to $\mathfrak A$. A family $\mathcal F$ of obstructions for $\mathfrak A$ is called a \emph{complete set of obstructions} if for every $\tau$-structure $\mathfrak B$ that does not admit a homomorphism to $\mathfrak A$ there exists some $\mathfrak C \in \mathcal F$ which admits a homomorphism to $\mathfrak B$. The
structure $\A$ is said to have \emph{finite duality} if it admits a finite complete set of obstructions. 
An obstruction $\mathfrak C$ for $\mathfrak A$ is called \emph{critical} if every proper (not necessarily induced) substructure of $\mathfrak C$ admits a homomorphism to $\mathfrak A$. For any set $A$, let $\mathrm{pr}^n_k$ denote the projection map from
$A^n$ to $A$ which maps any tuple to its $k$-th coordinate. We claim the following (essentially from \cite{LLT}:

\textbf{Claim 2.} If there exists an $(n+1)$-ary $1$-tolerant polymorphism of $\mathfrak A$ then the critical obstructions of $\mathfrak A$ 
have at most $n$ hyperedges. If $\mathfrak A$ is a monster extension, the converse holds as well.

\textit{Proof of Claim 2.} (Forwards.) By contraposition.
Let $\mathfrak C$ be a critical obstruction of $\mathfrak A$
with $m$ distinct hyperedges $e_1,\dots,e_m$, $m > n$. Then
for $k \in \{1,\dots,m\}$, the $\tau$-structure $\mathfrak C_k$ obtained from $\mathfrak C$ by removing $e_k$
(without changing the domain) admits a homomorphism $h_k$
to $\mathfrak A$. By definition of $^1\mathfrak A^m$, the map
$h=(h_1,\dots,h_m)$ is a homomorphism from $\mathfrak C$ to $^1\mathfrak A^m$. Therefore there is no homomorphism from $^1\mathfrak A^m$ to $\mathfrak A$, and in particular none from $^1\mathfrak A^{n+1}$ to $\mathfrak A$.

(Backwards.)
Conversely, suppose that $\mathfrak A$ is a monster extension, and that there is no homomorphism from $^1\mathfrak A^{n+1}$ to $\mathfrak A$. It follows from Claim 1 that
there exists a finite substructure $\mathfrak C$ of $^1\mathfrak A^{n+1}$ which has no homomorphism to $\mathfrak A$. 
Hence, $\mathfrak C$ is an obstruction of $\mathfrak A$ which admits a homomorphism $h$ to $^1\mathfrak A^{n+1}$. 
Let $\mathfrak C'$ be a (not necessarily induced) substructure of $\mathfrak C$ that
is critical (such a $\mathfrak C'$ always exists).
For every $k \in \{1,\dots,n+1\}$ there exists a hyperedge $e_k$
of $\mathfrak C'$ which is not preserved by $\mathrm{pr}_k^{n+1} \circ h$,
since $\mathrm{pr}_k^{n+1} \circ h$ is not a homomorphism from $\mathfrak C$ to $\mathfrak A$. By definition of $^1\mathfrak A^{n+1}$,
$e_k$ is respected by $\mathrm{pr}_j^{n+1} \circ h$ for every $j \neq k$, and thus $e_j \neq e_k$ for $j \neq k$. Therefore $\mathfrak C$ has at least $n+1$ hyperedges. $\Box$

\textit{Proof of Theorem.} (Forwards.)
Suppose first that CSP$(\mathfrak A)$ is first-order definable. 
Since CSP$(\mathfrak A)$ is a class of finite structures that is
closed under inverse homomorphisms, by the dual version of Rossman's Theorem, \cite{Rossman08}, there is a universal negative first-order
$\tau$-sentence $\phi$ that holds on a finite structure $\mathfrak B$ if and only if $\mathfrak B$ homomorphically maps to $\mathfrak A$. Bringing $\phi$ into prenex negation normal form, it is straightforward to read from $\phi$ a finite complete set $\mathcal F$ of obstructions to $\mathfrak A$. Let $m$ be the maximal number of hyperedges in the obstructions from $\mathcal F$. 
By the claim above, since $\mathfrak A$ is a monster, there is a homomorphism from
$^1\mathfrak A^{m+1}$ to $\mathfrak A$. This is by definition
a $1$-tolerant polymorphism of $\mathfrak A$.

(Backwards.) Now suppose that for some $n$ the structure $^1\mathfrak A^{n+1}$
admits a homomorphism to $\mathfrak A$. By the claim above the critical obstructions of $\mathfrak A$ have at most $n$ hyperedges. Since our signature is finite and relational,
this implies that there are finitely many critical obstructions to $\mathfrak A$. This
implies that the set of all critical obstructions is a finite obstruction
set for $\mathfrak A$. It is now straightforward to write down
a (universal) first-order definition of CSP$(\mathfrak A)$.
\qed
\begin{cor}
Let $\mathfrak A$ be a structure on a finite signature.
Then CSP$(\mathfrak A)$ is first-order definable if and only if $\mathfrak A$ has an elementary extension which has a $1$-tolerant polymorphism.
\end{cor}
\proof
By Lemma~\ref{lem:monster-model}, $\mathfrak A$ has a monster elementary extension $\M$.
Since $\mathfrak M$ and $\mathfrak A$ satisfy the same primitive positive sentences, CSP$(\mathfrak A)$ is
first-order definable if and only if CSP$(\mathfrak M)$ is.
The statement follows immediately from
Theorem~\ref{thm:fo}.
\qed

\subsection{Horn definability}

We will briefly examine a class of structures for which we can give a neat algebraic condition as to whether a relation that is qf-definable admits a qf Horn definition. Recalling known complexity results for fo-expansions of $(\mathbb{R};+,1)$ we will see that the presence of a certain polymorphism exactly delineates those fo-expansions for which the CSP is NP-complete from those which are in P. The following proposition is essentially from \cite{Maximal}.
\begin{prop}
\label{prop:Horn}
Let $\A$ be a structure with an embedding $e$ from $\A^2$ into $\A$. Then a relation $R$ that is qf-definable in the relations of $\A$ is preserved by $e$ iff it admits a qf-Horn definition in $\A$.
\end{prop}
\proof
In this proof $\overline{x}_1\ldots,\overline{x}_k$ should be read as variable subtuples of the variable tuple $\overline{x}$. Likewise with the element subtuples $\overline{a}_1\ldots,\overline{a}_k$ of $\overline{a}$.

(Backwards.) Let $\mathcal{F}$ be a Horn definition of $R$.  Suppose $\overline{a}$ and $\overline{a}' \in R^\A$. It suffices to demonstrate the preservation of each clause in $\mathcal{F}$ of the form $(R_1(\overline{x}_1) \wedge \ldots \wedge R_l(\overline{x}_l)) \rightarrow R_{l+1}(\overline{x}_{l+1})$, for $R_1,\ldots,R_{l+1}$ relations of $\A$.
\[
\begin{array}{ccccccc}
(R_1(\overline{a}_1) & \wedge & \ldots & \wedge & R_l(\overline{a}_l)) & \rightarrow & R_{l+1}(\overline{a}_{l+1}) \\
(R_1(\overline{a}'_1) & \wedge & \ldots & \wedge & R_l(\overline{a}'_l)) & \rightarrow & R_{l+1}(\overline{a}'_{l+1}) \\
\hline
(R_1(e(\overline{a}_1,\overline{a}'_1)) & \wedge & \ldots & \wedge & R_l(e(\overline{a}_l,\overline{a}'_l))) & \rightarrow & R_{l+1}(e(\overline{a}_{l+1},\overline{a}'_{l+1})) \\
\end{array}
\]
If the former clauses are true, there are two cases. Either some antecedent $R_i(\overline{a}_i)$ or $R_i(\overline{a}'_i)$ is false, in which case  $R_i(e(\overline{a}_i,\overline{a}'_i))$ is false, and the latter clause is true. Or, if all antecedents in both former clauses are true, then both $R_{l+1}(\overline{a}_{l+1})$ and $R_{l+1}(\overline{a}'_{l+1})$ are true, so it follows that $R_{l+1}(e(\overline{a}_{l+1},\overline{a}'_{l+1}))$ is true, and the latter clause is true.

(Forwards.) By contraposition. Consider a CNF definition $\mathcal{F}$ of $R$ in $\A$ that is irreducible in the sense that it has no redundant literals in its clauses. If it is not Horn, there exists a clause $R_1(\overline{x}_1) \vee R_2(\overline{x}_2) \vee S_3(\overline{x}_3) \vee \ldots \vee S_{l}(\overline{x}_l)$, with $R_1,R_2$ positive literals $S_3,\ldots,S_l$ positive or negative literals,  with $\overline{a},\overline{a}' \in R^\A$ s.t. 
\[
\begin{array}{c}
R_1(\overline{a}_1) \wedge \neg R_2(\overline{a}_2) \wedge \neg S_3(\overline{a}_3) \wedge \ldots \wedge \neg S_l(\overline{a}_l) \\
\neg R_1(\overline{a}'_1) \wedge R_2(\overline{a}'_2) \wedge \neg S_3(\overline{a}'_3) \wedge \ldots \wedge \neg S_l(\overline{a}'_l) \\
\end{array}
\]
Consider the tuple $e(\overline{a},\overline{a}')$. Clearly it will fail to satisfy the clause.
\qed
\noindent We have already met an example of a structure with such an embedding: $(\mathbb{R};+,1)$.
\begin{cor}
Let $\B$ be an fo-expansion of $(\mathbb{R};+,1)$ and let $e:(\mathbb{R};+,1)^2\rightarrow (\mathbb{R};+,1)$ be an embedding. Then: if $e$ is a polymorphism of $\B$, then CSP($\B$) is in P; otherwise CSP($\B$) is NP-complete.
\end{cor}
\proof
Note that $(\mathbb{R};+,1)$ admits quantifier elimination and so all fo-expansions may be specified as qf CNFs. It is proved in \cite{HornOrFull} that those that admit qf-Horn definitions give a CSP that is in P while those that do not give CSPs that are NP-complete. The result follows from Proposition~\ref{prop:Horn}.
\qed

\section{Concluding remarks and open problems}
The universal-algebraic approach to the complexity of CSPs
relies on two basic facts: the fact that every finite or $\omega$-categorical structure is homomorphically equivalent to a model-complete core structure, and that
primitive positive definability is characterized by preservation under polymorphisms.

In this paper we have presented generalizations of those two facts to not necessarily
$\omega$-categorical structures. A key concept in our proofs is the concept of
existential-positive closure, which we used to re-derive and generalize results in the literature about the presence and uniqueness of model-complete cores \cite{Cores-journal}.
We also give a necessary and sufficient condition as to when a CSP can be formulated with an $\omega$-categorical template.

The results of this paper show that the second part of the universal algebraic approach --
the characterization of pp-definability by polymorphisms -- can be applied to study the complexity of CSP$(\mathfrak A)$ for arbitrary (and not just $\omega$-categorical) infinite-domain structures $\mathfrak A$. Among one of the first applications, we
\begin{iteMize}{$\bullet$}
\item gave a polymorphism-based characterization of those CSPs that are fo-definable;
\item demonstrated in the context of real-valued constraint satisfaction problems that for large classes of CSPs the border between easy and hard constraint
satisfaction can be described in terms of the existence of a certain polymorphism of the constraint language;
\item have presented a strong universal-algebraic hardness criterion
based on the absence of essential polymorphisms from the constraint language.
\end{iteMize}

The following question is left for future research: can we strengthen our preservation theorem (Theorem~\ref{thm:inv-pol-inf}) to show, under the additional assumption that $\mathfrak A$ is epc (or epc and saturated, or has a model-complete core theory), that a first-order definable relation is pp-definable if and only if it is preserved by the \emph{finitary} polymorphisms of $\mathfrak A$?

\section*{Acknowledgements}

We are grateful to our referees. Moreover, we would like to thank Moritz M\"uller for numerous comments which helped to improve the paper.

\bibliographystyle{acm}
\bibliography{global}

\begin{thebibliography}{10}

\bibitem{BenYaacov}
{\sc {Ben Yaacov}, I.}
\newblock Positive model theory and compact abstract theories.
\newblock {\em Journal of Mathematical Logic 3}, 1 (2003), 85--118.

\bibitem{Bodirsky}
{\sc Bodirsky, M.}
\newblock Constraint satisfaction with infinite domains.
\newblock Dissertation an der Humboldt-Universit\"at zu Berlin, 2004.

\bibitem{Cores-journal}
{\sc Bodirsky, M.}
\newblock Cores of countably categorical structures.
\newblock {\em Logical Methods in Computer Science 3}, 1 (2007), 1--16.

\bibitem{Maximal}
{\sc Bodirsky, M., Chen, H., K\'ara, J., and von Oertzen, T.}
\newblock Maximal infinite-valued constraint languages.
\newblock {\em Theoretical Computer Science (TCS) 410\/} (2009), 1684--1693.
\newblock A preliminary version appeared at ICALP'07.

\bibitem{BodirskyGrohe}
{\sc Bodirsky, M., and Grohe, M.}
\newblock Non-dichotomies in constraint satisfaction complexity.
\newblock In {\em Proceedings of ICALP'08\/} (2008), pp.~184--196.

\bibitem{BodirskyHermannRichoux}
{\sc Bodirsky, M., Hermann, M., and Richoux, F.}
\newblock Complexity of existential positive first-order logic.
\newblock In {\em Proceedings of Computing in Europe (CiE'09)\/} (2009),
  pp.~31--36.

\bibitem{BodHilsMartin}
{\sc Bodirsky, M., Hils, M., and Martin, B.}
\newblock On the scope of the universal-algebraic approach to constraint
  satisfaction.
\newblock In {\em Proceedings of LICS'10\/} (2010).

\bibitem{HornOrFull}
{\sc Bodirsky, M., Jonsson, P., and von Oertzen, T.}
\newblock Horn versus full first-order: a complexity dichotomy for algebraic
  constraint satisfaction problems.
\newblock {\em To appear in the Journal of Logic and Computation\/} (2010).

\bibitem{ecsps}
{\sc Bodirsky, M., and K\'ara, J.}
\newblock The complexity of equality constraint languages.
\newblock {\em Theory of Computing Systems 3}, 2 (2008), 136--158.
\newblock A conference version appeared in the proceedings of CSR'06.

\bibitem{tcsps}
{\sc Bodirsky, M., and K\'ara, J.}
\newblock The complexity of temporal constraint satisfaction problems.
\newblock In {\em Proceedings of the Symposium on Theory of Computing (STOC)\/}
  (2008), pp.~29--38.

\bibitem{BodirskyNesetrilJLC}
{\sc Bodirsky, M., and Ne\v{s}et\v{r}il, J.}
\newblock Constraint satisfaction with countable homogeneous templates.
\newblock {\em Journal of Logic and Computation 16}, 3 (2006), 359--373.

\bibitem{BoKaKoRo}
{\sc Bodnar\v{c}uk, V.~G., Kalu\v{z}nin, L.~A., Kotov, V.~N., and Romov, B.~A.}
\newblock Galois theory for post algebras, part {I} and {II}.
\newblock {\em Cybernetics 5\/} (1969), 243--539.

\bibitem{Conservative}
{\sc Bulatov, A.}
\newblock Tractable conservative constraint satisfaction problems.
\newblock In {\em Proceedings of LICS'03\/} (2003), pp.~321--330.

\bibitem{Bulatov}
{\sc Bulatov, A.}
\newblock A dichotomy theorem for constraint satisfaction problems on a
  3-element set.
\newblock {\em Journal of the ACM 53}, 1 (2006), 66--120.

\bibitem{JBK}
{\sc Bulatov, A., Krokhin, A., and Jeavons, P.~G.}
\newblock Classifying the complexity of constraints using finite algebras.
\newblock {\em SIAM Journal on Computing 34\/} (2005), 720--742.

\bibitem{DBLP:conf/dagstuhl/BulatovV08}
{\sc Bulatov, A.~A., and Valeriote, M.}
\newblock Recent results on the algebraic approach to the {CSP}.
\newblock In {\em Complexity of Constraints\/} (2008), pp.~68--92.

\bibitem{CSPSurveys}
{\sc Creignou, N., Kolaitis, P.~G., and Vollmer, H.}, Eds.
\newblock {\em Complexity of Constraints - An Overview of Current Research
  Themes [Result of a Dagstuhl Seminar]\/} (2008), vol.~5250 of {\em Lecture
  Notes in Computer Science}, Springer.

\bibitem{FederVardi}
{\sc Feder, T., and Vardi, M.}
\newblock The computational structure of monotone monadic {SNP} and constraint
  satisfaction: {A} study through {D}atalog and group theory.
\newblock {\em {SIAM} Journal on Computing 28\/} (1999), 57--104.

\bibitem{Geiger}
{\sc Geiger, D.}
\newblock Closed systems of functions and predicates.
\newblock {\em Pacific Journal of Mathematics 27\/} (1968), 95--100.

\bibitem{HobbyMcKenzie}
{\sc Hobby, D., and McKenzie, R.}
\newblock {\em The Structure of Finite Algebras}.
\newblock AMS, 1988.

\bibitem{HodgesLong}
{\sc Hodges, W.}
\newblock {\em Model theory}.
\newblock Cambridge University Press, 1993.

\bibitem{Hodges}
{\sc Hodges, W.}
\newblock {\em A shorter model theory}.
\newblock Cambridge University Press, Cambridge, 1997.

\bibitem{JeavonsClosure}
{\sc Jeavons, P., Cohen, D., and Gyssens, M.}
\newblock Closure properties of constraints.
\newblock {\em Journal of the ACM 44}, 4 (1997), 527--548.

\bibitem{Jeavons}
{\sc Jeavons, P.~G.}
\newblock On the algebraic structure of combinatorial problems.
\newblock {\em Theoretical Computer Science 200\/} (1998), 185--204.

\bibitem{Ladner}
{\sc Ladner, R.~E.}
\newblock On the structure of polynomial time reducibility.
\newblock {\em Journal of the ACM 22}, 1 (1975), 155--171.

\bibitem{LLT}
{\sc Larose, B., Loten, C., and Tardif, C.}
\newblock A characterisation of first-order constraint satisfaction problems.
\newblock {\em Logical Methods in Computer Science, DOI: 10.2168/LMCS-3(4:6)\/}
  (2007).

\bibitem{Poizat:AlgebresDePost}
{\sc Poizat, B.}
\newblock Theorie de {Galois} pour les algebres de {Post} infinitaires.
\newblock {\em Z. Math. Logik Grudl. Math. 27\/} (1981), 31--44.

\bibitem{Reingold}
{\sc Reingold, O.}
\newblock Undirected connectivity in log-space.
\newblock {\em Journal of the ACM 55}, 4 (2008).

\bibitem{RomovPosPrimStruc}
{\sc Romov, B.}
\newblock Positive primitive structures.
\newblock In {\em Multiple-Valued Logic, 2009. ISMVL '09. 39th International
  Symposium on\/} (may 2009), pp.~72 --76.

\bibitem{RomovISMVL04}
{\sc Romov, B.~A.}
\newblock Some properties of local partial clones on an infinite set.
\newblock In {\em ISMVL '04: Proceedings of the 34th International Symposium on
  Multiple-Valued Logic\/} (Washington, DC, USA, 2004), IEEE Computer Society,
  pp.~120--125.

\bibitem{Rossman08}
{\sc Rossman, B.}
\newblock Homomorphism preservation theorems.
\newblock {\em J. ACM 55}, 3 (2008).

\bibitem{Simmons72}
{\sc Simmons, H.}
\newblock Existentially closed structures.
\newblock {\em J. Symb. Log. 37}, 2 (1972), 293--310.

\bibitem{Simmons76}
{\sc Simmons, H.}
\newblock Large and small existentially closed structures.
\newblock {\em Journal of Symbolic Logic 41}, 2 (1976), 379--390.

\bibitem{SzaboLoc}
{\sc Szab\'{o}, L.}
\newblock Concrete representation of relational structures of universal
  algebras.
\newblock {\em Acta Sci. Math. (Szeged) 40\/} (1978), 175--184.

\bibitem{Szendrei}
{\sc Szendrei, A.}
\newblock {\em Clones in universal Algebra}.
\newblock S\'eminaire de Math\'ematiques Sup\'erieures. Les Presses de
  L'Universit\'e de {M}ontr\'eal, 1986.

\bibitem{Zilber}
{\sc Zilber, B.}
\newblock {\em Uncountable categorical theories, Tranlations of Mathematical
  Monographs}, vol.~117.
\newblock Amer. Math. Soc., 1993.

\end{thebibliography}

\end{document}